\documentclass[12pt]{article} \textheight= 220mm \textwidth=160mm \topmargin = -1 cm
\usepackage{makeidx}
\usepackage{graphicx}
\usepackage{amssymb}
\input epsf
\newcommand{\D}{{\rm d}}

\newcommand{\LL}{\mbox{\boldmath$L$}}

\newcommand{\MM}{\mbox{\boldmath$M$}}
\newcommand{\NN}{\mbox{\boldmath$N$}}

\newcommand{\QQ}{\mbox{\boldmath$Q$}}

\newtheorem{theorem}{Theorem}

\newtheorem{lemma}{Lemma}

\newtheorem{proposition}{Proposition}

\newtheorem{definition}{Definition}

\begin{document}

\title{Selection theorem for systems with inheritance}

\author{Alexander N. Gorban\\
University of Leicester \\ LE1 7RH Leicester, UK; \\ e-mail:
ag153@le.ac.uk}

\date{}

\maketitle

\begin{abstract}
The problem of finite-dimensional asymptotics of
infinite-dimensional dynamic systems is studied. A non-linear
kinetic system with {\it conservation of supports} for
distributions has generically finite-dimensional asymptotics. Such
systems are apparent in many areas of biology, physics (the theory
of parametric wave interaction), chemistry and economics. This
conservation of support has a biological interpretation: {\it
inheritance}. The finite-dimensional asymptotics demonstrates
effects of {\it ``natural" selection}. Estimations of the
asymptotic dimension are presented. After some initial time,
solution of a kinetic equation with conservation of support
becomes a finite set of narrow peaks that become increasingly
narrow over time and move increasingly slowly. It is possible that
these peaks do not tend to fixed positions, and the path covered
tends to infinity as $t \rightarrow \infty$. The {\it drift
equations} for peak motion are obtained. Various types of
distribution stability are studied: internal stability (stability
with respect to perturbations that do not extend the support),
external stability or uninvadability (stability with respect to
strongly small perturbations that extend the support), and stable
realizability (stability with respect to small shifts and
extensions of the density peaks). Models of self-synchronization
of cell division are studied, as an example of selection in
systems with additional symmetry. Appropriate construction of the
notion of typicalness in infinite-dimensional space is discussed,
and the notion of ``completely thin" sets is introduced.

\end{abstract}

\clearpage

\tableofcontents

\clearpage

\section{Introduction: Unusual conservation law}

There are three geometrically distinguished classes of dynamical
systems originating from natural sciences:
\begin{itemize}
\item{Hamiltonian systems;}
\item{Dissipative systems with entropy (or another thermodynamic Lyapunov function); and}
\item{Systems with inheritance.}
\end{itemize}
Hamiltonian systems originated from mechanics. Symplectic geometry
followed them. Dissipative systems with thermodynamic Lyapunov
functions arose from thermodynamics and kinetics. The related
geometry is the geometry of Legendre transformation and contact
structures (this subject is not exhausted yet; see for example
\cite{Yab,G1} and a recent publication \cite{GRMEL}). Systems with
inheritance are emerging from population dynamics, physical
kinetics, turbulence theory, and many other fields of natural
science. The geometrical sense of inheritance is a special
conservation law, in which the conserved ``quantity" is a set, a
support of distribution.

In the 1970s to the 1980s, theoretical work developed another
``common" field simultaneously applicable to physics, biology and
mathematics. For physics it is (so far) part of the theory of a
special kind of approximation, demonstrating, in particular,
interesting mechanisms of discreteness in the course of the
evolution of distributions with initially smooth densities. However,
what for physics is merely a convenient approximation is a
fundamental law in biology: inheritance. The consequences of
inheritance (collected in the selection theory
\cite{Darwin,Haldane,Mayr,Maynard73,Maynard82,Bish,Ewens,Hammerstein,G1,GorKhleb,SemSem})
give one of the most important tools for biological reasoning.

This paper is not a review of the scientific literature on
evolution, and we mention here only references that are particularly
important for our understanding of the selection theory and its
applications.

Consider a community of animals. Let it be biologically isolated.
Mutations can be neglected in the first approximation. In this case,
new genes do not emerge.

An example from physics is as follows. Let waves with wave vectors
$k$ be excited in some system. Denote $K$ a set of wave vectors $k$
of excited waves. Let the wave interaction does not lead to the
generation of waves with new $k \notin K$. Such an approximation is
applicable to a variety of situations, and has been described in
detail for wave turbulence in \cite{Zakharov1,Zakharov2}.

What is common in these examples is the evolution of a distribution
with a support that does not increase over time.

{\it What does not increase must, as a rule, decrease, if the
decrease is not prohibited.} This naive thesis can be converted into
rigorous theorems for the case under consideration \cite{G1}. It is
proved that the support decreases in the limit $t\rightarrow \infty$
if it was sufficiently large initially. (At finite times the
distribution supports are conserved and decrease only in the limit
$t\rightarrow \infty$.) Conservation of the support usually results
in the following effect: dynamics of an initially
infinite-dimensional system at $t\rightarrow \infty$ can be
described by finite-dimensional systems.

The simplest and most common class of equations in applications for
which the distribution support does not grow over time is
constructed as follows. To each distribution $\mu$ is assigned a
function $k_{\mu}$ by which distributions can be multiplied. Let us
write down the equation:
\begin{equation} \label{sel1}
{\D \mu \over \D t}= k_{\mu} \times \mu.
\end{equation}

The multiplier $k_{\mu}$ is called a {\it reproduction coefficient}.
It depends on $\mu$, and this dependence can be rather general and
non-linear.

Two remarks can be important:
\begin{enumerate}
\item{The apparently simple form of (\ref{sel1}) does not mean
that this system is linear or even close to linear. The operator
$\mu \mapsto k_{\mu}$ is a general non-linear operator, and the only
restriction is its continuity in an appropriate sense (see below).}
\item{On a finite set $X=\{x_1, \ldots, x_n \}$, non-negative measures $\mu$
are simply non-negative vectors $\mu_i \geq 0$ (i=1,\ldots, n), and
(\ref{sel1}) appears to be a system of equations of the following
type:
\begin{equation} \label{selfin}
{\D \mu_i \over \D t}= k_i(\mu_1, \ldots, \mu_n) \times \mu_i,
\end{equation}
and the only difference from a general dynamic system is the special
behavior of the right-hand side of (\ref{selfin}) near zero values
of $\mu_i$.}
\end{enumerate}

The right-hand side of (\ref{sel1}) is the product of the function
$k_{\mu}$ and the distribution $\mu$, and hence $ \D \mu /  \D t$
should be zero when $\mu$ is equal to zero; therefore the support of
$\mu$ is conserved in time (over finite times).

Let us start a more formal consideration, with basic definitions and
notations. First, we introduce the {\it space of inherited units}
$X$. In this paper $X$ is a compact metric space with a metric
$\rho(x,y)$. In other special sections we assume that $X$ is a
closed bounded domain in finite-dimensional real space $R^n$. As a
particular case of compact space, a finite set $X$ can be discussed.

We study the dynamics of distributions on $X$. Each distribution on
a compact space $X$ is a continuous linear functional on the space
of continuous real functions $C(X)$. We follow the Bourbaki approach
\cite{Bourbaki}: a measure is a continuous functional, an integral.
Bourbaki's book \cite{Bourbaki} contains all the necessary notions
and theorems (and much more material than we need here). Space
$C(X)$ is a Banach space endowed with the maximum norm
\begin{equation} \label{uninorm}
\|f\| = \max_{x \in X} |f(x)|.
\end{equation}
If $\mu \in C^*(X)$ is a continuous function and $f \in C(X)$, then
$[\mu,f]$ is the value of $\mu$ at a function $f$.

Let us mention here two other notations. If $X$ is a bounded closed
subset of a finite-dimensional space $R^n$, then we represent this
functional as the integral
\begin{equation}\label{distribut}
[\mu,f] = \int \mu(x) f(x) \,  \D x,
\end{equation}
which is the standard notation for distribution (or generalized
function) theory. Note that here the ``density" $\mu(x)$ is not
assumed to be an absolute continuous function with respect to the
Lebesgue measure $\D x$ (or even a function), and the notation in
Eq. (\ref{distribut}) has the same sense as $[\mu,f]$. If the
measure is defined as a function on a $\sigma$-algebra of sets, then
the following notation is used:
\begin{equation}\label{meastheor}
[\mu,f] = \int  f(x) \,  \mu(\D x).
\end{equation}

We use the notation $[\mu,f]$ for general spaces $X$ and the
representation (\ref{distribut}) on domains in $R^n$ without any
additional comments. The product $k \times \mu$ is defined for any
$k \in C(X)$, $\mu \in C^*(X)$ by the equality: $[k
\mu,f]=[\mu,kf]$.

The support of $\mu$, ${\rm supp} \mu$, is the smallest closed
subset of $X$ with the following property: if $f(x)= 0$ on ${\rm
supp} \mu$, then $[\mu,f]=0$, i.e. $\mu(x)=0$ outside ${\rm supp}
\mu$.

In the space of measures we use {\it weak$^*$ convergence}, i.e. the
convergence of averages:
\begin{equation} \label{weak}
\mu_i \rightarrow \mu^* \; \mbox{if and only if} \; [\mu_i, \varphi]
\rightarrow [ \mu^*, \varphi]
\end{equation}
for all continuous functions $\varphi \in C(X)$. This weak$^*$
convergence of measures generates {\it weak$^*$ topology} on the
space of measures (sometimes called weak topology of conjugated
space, or wide topology).

{\it Strong topology} on the space of measures $C^*(X)$ is defined
by the norm $\|\mu\|=sup_{\|f\|=1}[\mu,f]$.

Strictly speaking, the space on which $\mu$ is defined and the
distribution class it belongs to should be specified. The properties
of the mapping $\mu \mapsto k_{\mu}$ should also be specified, and
the existence and uniqueness of solutions of (\ref{sel1}) under
given initial conditions should be identified. In specific
situations the answers to these questions are not difficult.

The sequence of continuous functions $k_i(x)$ is considered to be
convergent if it converges uniformly. The sequence of measures
$\mu_i$ is called convergent if for any continuous function
$\varphi$ the integrals $[\mu_i,\varphi]$ converge [weak$^*$
convergence (\ref{weak})]. The mapping $\mu\mapsto k_{\mu}$
assigning the reproduction coefficient $k_{\mu}$ to the measure
$\mu$ is assumed to be continuous with respect to these
convergencies.

Finally, the space of measures is assumed to have a bounded set
$M$ that is positively invariant relative to system (\ref{sel1}):
if $\mu(0) \in M$, then $\mu(t) \in M$ (we also assume that $M$ is
non-trivial, i.e. it is neither empty nor a one-point set). This
$M$ serves as the phase space of system (\ref{sel1}). (Let us
remind that the set of measures $M$ is bounded if the set of
integrals $\{[\mu,f] \, | \, \mu \in M, \|f\|\leq1\}$ is bounded,
where $\|f\|$ is the norm (\ref{uninorm}).) We study dynamic of
system (\ref{sel1}) in bounded positively invariant set $M$.

Most of the results for systems with inheritance use a {\bf
theorem on weak$^*$ compactness}: {\it The bounded set of measures
is precompact with respect to weak$^*$ convergence (i.e. its
closure is compact).} Therefore the set of corresponding
reproduction coefficients $k_M = \{k_{\mu}\, | \,\mu \in M \}$ is
precompact.

Let us start with the simplest example and the first theorem, and
then discuss possible interpretations. The simplest example of an
emerging discrete distribution from a continuous initial
distribution gives us the following equation:
\begin{equation} \label{sel2}
{\partial \mu (x,t) \over \partial t}= \left[f_0(x)- \int_a^b
f_1(x)\mu(x,t) \,  \D x \right] \mu(x,t),
\end{equation}
where the functions $f_0(x)$ and $f_1(x)$ are positive and
continuous on the closed segment $[a,b]$. Let the function $f_0(x)$
reach the global maximum on the segment $[a,b]$ at a single point
$x_0$. If $x_0 \in {\rm supp} \mu(x,0)$, then:
\begin{equation} \label{sel3}
\mu(x,t) \rightarrow {f_0(x_0) \over f_1(x_0)}\delta(x-x_0), \;
\quad \quad \mbox{when} \; t\rightarrow\infty,
\end{equation}
where $\delta(x-x_0)$ is the $\delta$-function.

If $f_0(x)$ has several global maxima, then the right-hand side of
(\ref{sel3}) can be the sum of a finite number of
$\delta$-functions. Here a natural question arises: is it worth
considering such a possibility? Should not we deem it improbable for
$f_0(x)$ to have more than one global maximum? Indeed, such a case
seems to be very unlikely to occur. More details on this are given
below.

The limit behavior of a typical system with inheritance (\ref{sel1})
can be much more complicated than (\ref{sel3}). Here we can mention
that any finite-dimensional system with a compact phase space can be
embedded in a system with inheritance (\ref{selfin}). An additional
possibility for the limit behavior is, for example, the drift effect
(Section~\ref{drift}).

The first step in the routine investigation of a dynamical system is
a question about fixed points and their stability. The first
observation concerning the system  (\ref{sel1}) is that it can only
be asymptotically stable for steady-state distributions, the support
of which is discrete (i.e. the sums of $\delta$-functions). This can
be proved for all consistent formalizations and can be understood as
follows.

Let $U$ be a domain in $X$, and the ``total amount" in $U$
(integral of $|\mu|$ over $U$) be less than $\varepsilon>0$ but
not equal to zero. Let us substitute distribution $\mu$ by zero on
$U$ (the rest remains as it is). It is natural to consider this
disturbance of $\mu$ as $\varepsilon$-small. However, if the
dynamics is described by (\ref{sel1}), there is no way back to the
undisturbed distribution, because the support cannot increase. If
the steady-state distribution $\mu^*$ is asymptotically stable,
then for some $\varepsilon>0$ any $\varepsilon$-small perturbation
of $\mu^*$ relaxes back to $\mu^*$. This is possible only if for
any domain $U$ the integral of $|\mu^*|$ over $U$ is either $0$ or
greater than $\varepsilon$. Hence, this asymptotically stable
distribution $\mu^*$ is the sum of a finite number of point
measures:
\begin{equation} \label{asydelta}
\mu^*(x)=\sum_{i=1}^qN_i \delta_{x_i}
\end{equation}
with $|N_i|>\varepsilon$ for all $i$ and where $\delta_{x_i}$ is
the normalized point measure at point $x_i$. In distribution
theory notation, this corresponds to the $\delta$-function
$\delta(x-x_i)$.

The perturbation discussed is small not only in the weak$^*$
topology, but also in the strong sense, and thus it is sufficient to
consider strongly small perturbations to prove that the
asymptotically stable distribution should be discrete. Hence, this
statement is true if the operator $\mu \mapsto k_{\mu}$ is
continuous for strong topology on the space of measures. This is a
significantly weaker requirement than being continuous in weak$^*$
topology.

Thus, we have the first theorem.
\begin{theorem}
The support of asymptotically stable distributions for the system
(\ref{sel1}) is always discrete.
\end{theorem}
This simple observation has many strong generalizations to general
$\omega$-limit points, to equations for vector measures, etc.

Dynamic systems in which the phase variable is a distribution $\mu$
and distribution support is the integral of motion frequently occur
in both physics and biology. Because of their attractive properties,
they are frequently used as approximations: we try to find the
``main part" of the system in the form of (\ref{sel1}), and
represent the rest as a small perturbation of the main part.

Equations in the form of (\ref{sel1}) allow the following biological
interpretation: $\mu$ is the distribution of the number (or of
biomass or another extensive variable) over inherited units:
species, varieties, supergenes, genes. Whatever is considered as the
{\it inherited unit} depends on the context and the specific
problem. The value of $k_{\mu}(x)$ is the reproduction coefficient
of the {\it inherited unit} $x$ under given conditions. The notion
of ``given conditions" includes the distribution $\mu$, and the
reproduction coefficient depends on $\mu$. Equation (\ref{sel2}) can
be interpreted as follows: if $f_0(x)$ is the specific birth-rate of
the inherited unit $x$ (below, for the sake of definiteness, $x$ is
a variety, following the spirit of Darwin's seminal work
\cite{Darwin}), the death rate for the representatives of all
inherited units (varieties) is determined by one common factor
depending on the density $\int_a^b f_1(x)\mu(x,t) \, \D x $;
$f_1(x)$ is the individual contribution of variety $x$ to this death
rate.

On the other hand, for systems of waves with a parametric
interaction, $k_{\mu}(x)$ can be the amplification (decay) rate of
the wave with wave vector $x$.

Conservation of the support in (\ref{sel1}) can be considered as
inheritance, and we call system (\ref{sel1}) and its nearest
generalizations ``systems with inheritance". Traditional
separation of the process of transferring biological information
into inheritance and mutations, which are small in any admissible
sense, can be compared to a description according to the following
pattern: system (\ref{sel1}) (or its nearest generalizations) plus
small disturbances. Beyond the limits of such a description,
discussion of inheritance loses the conventional sense.

In biology such an approximation is essentially applicable to all
classical genetics, and to the formal content of the theory of
natural selection. The initial diversity is ``thinned out" over
time, and the limit distribution supports are described by some
extremal principles (principles of optimality).

The first study of dynamics systems with inheritance was carried out
by J.B.S. Haldane. He used the simplest examples, studied
steady-state distributions and obtained the extremal principle for
them. His pioneering book ``The Causes of Evolution" (1932)
\cite{Haldane} gives a clear explanation of the connections between
inheritance (the conservation of distribution support) and the
optimality of selected varieties.

Haldane's work was followed by an entirely independent series of
works on the S-approxi\-mation in the spin wave theory and on wave
turbulence \cite{Zakharov1,Zakharov2,Lvov}, which studied wave
configurations in the approximation of an ``inherited" wave vector,
and by ``synergetics" \cite{Synergetics}, in which the ``natural
selection" of modes is one of the basic concepts.

At the same time, a series of works on biological kinetics was
completed (see, for example \cite{RozSed,GorMik,SemSem,G1}). These
studies addressed not only steady states, but also general limit
distributions \cite{GorMik,G1} and waves in the space of inherited
units \cite{RozSed}. For steady states a new type of stability was
described -- stable realizability (see below). Many examples of
ecological applications are collected in reference \cite{SemSem}.
The application of optimality principles to crop growth simulation
is analyzed in reference \cite{crop}. Some attempts using
sociological applications are also known \cite{Sergo}.

The Haldane achievements were continued by works on stable
evolutionary strategies and evolutionary games. In works by Maynard
Smith \cite{Maynard73,Maynard82} the ``War of Attrition" model of
animal conflict was introduced and the notion of {\it evolutionarily
stable strategy} (ESS) was defined. This notion was elaborated
further in many papers
\cite{Taylor,Hofbauer,Thomas,VickCan,Cressman,Bom89,Bom90,Bom92,Bom95,Bom98,Bom02,Oechssler01,Oechssler02,EshSan}.
The reader is also referred to a recent review \cite{Hofbauer03}, in
which evolutionary game dynamics is defined as the application of
population dynamical methods to game theory. (It was invented by
evolutionary biologists, but had a great impact on modern game
theory and economical mathematics.) For some classes of models (a
``generalized war of attrition" \cite{Bish}), it was demonstrated
that either (i) there is no ESS or (ii) there is a unique ESS, which
is fully specified. In the case for which only a finite number of
pure strategies is available, global convergence to the ESS is
shown. Of course, for systems with inheritance (\ref{sel1}), more
complicated behavior is also possible. In reference \cite{VickCan}
collections of subsets that might be supports for ESSs were
identified. Imaginary experiments with mutant invasion are based on
the notion of EES \cite{Maynard73,Maynard82}. The dynamical
foundation of this notion and a dynamical theory of uninvadability
in the context of stability theory have been developed \cite{Bom90}.
It should be mentioned that the analogue of uninvadability, external
stability, was one of the main notions of the $S$-theory in physics
\cite{Zakharov1,Zakharov2,Lvov}, and ecological applications of this
notion were developed reasonably far in the 1970s--1980s; see
references \cite{G1,GorKhleb,SemSem} and references therein.

The dynamics of evolutionary games for the case of a continuum of
possible strategies has been investigated \cite{Bom95}. The
stability properties of stationary points were studied and some
examples were given. In fact, in reference \cite{Bom95} a particular
case of systems with inheritance was studied; in this case $X$ is
the space of strategies of an evolutionary game.

The setwise notions of stable evolutionary sets were introduced
\cite{Thomas} for evolutionary game models in which there is a
continuum of equilibrium states, with no state stable in itself, but
which together are evolutionarily stable. This concept was also
analyzed from a dynamical point of view \cite{Bom98}.

Recently, the theory of evolutionary games with a continuum of
possible strategies has been developed very intensively
\cite{Bom02,Oechssler01,Oechssler02,EshSan}; see also the review in
reference \cite{Hofbauer03}.

The first biological applications of systems with inheritance were
population dynamics and mathematical genetics. All the classical
equations for distributions of animals or genes have the form
(\ref{sel1}) (or a generalization with additional variables). The
space of inherited units, $X$, in these cases has a direct
biological interpretation: it is the space of inherited
variations, or species, or the space of alleles (``variations of
genes"). In ecological applications it has been proved
\cite{SemSem} that the concept of inherited variations of animals
(without consideration of alleles) gives appropriate accuracy in
the problem of succession, that is, in the modeling and simulation
of changes in a biological community under changing conditions.
But what is gene? Biology returns to this question again and again
\cite{gene?}. The interplay between ``units of function" and
``units of inheritance" for different time scale produces very
complicated and fascinating picture.

Epigenetic inherited units yield many interesting materials for
modeling. The source of dynamical difference between genetic and
epigenetic inheritance is their different time scales when they
are different \cite{Gene/Epi}. The interaction between these types
of inheritance could be quite mysterious. For example, the rates
and some specific properties of genetic mutation processes might
be inherited properties, as was discovered for the effect of
transgenerational instability \cite{Dubrova}. This phenomenon  is
probably due to an epigenetic mechanism.

For  the ``ecological time scale", the maternity effect forms
another group of inherited units. These units are important for
the evolution at ecological time-scales \cite{Matern}.

The space of inherited strategies provides the interpretation of $X$
in many applications. In particular, the selection of strategies of
the spatial distribution of individuals has been studied
\cite{Sad1}. In the case of non-monotonous dependence of the
reproduction coefficient on the mean population density, a cluster
formation was proven. This theory was applied to an investigation of
the creation of cellular clusters in flow-rate cultivators
\cite{Sad2}.

It is clear that animal migration is not completely random and that
it depends on conditions; in particular, predator migration depends
on space variations of the prey density, which might imply
interesting dynamical consequences, including changes in the number
and stability of equilibria and limit cycles \cite{Diek3}. Models of
evolutionarily optimal migration strategies in prey--predator
systems have been studied \cite{Sad3}. A great variety of dynamic
regimes has been observed, and some of them could be interpreted as
outbreak explosions.

The distribution of successors over time (that represent variations
of individual development, delayed maturation and even a pause in
ontogenesis) is important in the adaptation to stochastic
conditions. Evolutionarily optimal strategies of facultative
diapause for forest insects have been studied \cite{SemSem}. In
\cite{VanDor} a population with complicated dynamics was studied
numerically. It was demonstrated that random noise shifts the system
towards a higher probability of delaying maturation.

In ecological physiology, the points of $X$ represent strategies of
physiological adaptation. A useful notion is the adaptation
resource. The presentation of the adaptation process as a
redistribution of this resource for the neutralization of external
factors is an effective tool for adaptation modelling
\cite{GorPetMan}. These models of ``factors--resource" and the
dynamic theory of optimal evolutionary strategies allowed us to
develop ``correlation adaptometry" \cite{SedGorPet,GorManPer}. This
method of comparative ecological physiology is now in use for
comparative analysis of populations and groups for a wide range of
organisms, from the human population of the Far North
\cite{Pok,Svet} to herbaceous species \cite{Karman}.

The purpose of this paper is to present general results for the
theory of systems with inheritance: optimality principles for limit
distributions, theorems on selection, estimations of the limit
diversity (estimates of a number of points in the support for limit
distributions), the drift effect and drift equations. Some of these
results have been published in preprint \cite{GorMik} and,
partially, book form \cite{G1}.

The main benefit of the special form of systems with inheritance is
the possibility of describing the limit behavior of such systems by
avoiding the solution of equations. A system of weak and strong {\it
optimality principles} describes the supports of limit or stable
distributions. A special {\it drift asymptotic} reduces asymptotic
behavior at large time values by a finite system of ordinary
differential equations (ODEs). In subsequent sections this technique
is developed on the basis of investigation of the general system
represented by Equation (\ref{sel1}).

The outline of the paper is as follows. In the next section the
optimality principles for supports of $\omega$-limit distributions
are developed. These principles have a ``weak" form; the set of
possible supports is estimated from above and it is not obvious that
this estimation is effective (this is proved in
Section~\ref{selteor}).

Minimax estimations of the number of points in the support of
$\omega$-limit distributions are given in Section~\ref{estimates}.
The idea is to study systems under a $\varepsilon$-small
perturbation, to estimate the maximal number of points for each
realization of the perturbed system, and then to estimate the
minimum of these maxima among various realizations. These minimax
estimates can be constructive and do not use integration of the
system. The set of reproduction coefficients $\{k(\mu)\, | \, \mu
\in M \}$ is compact in $C(X)$. Therefore, this set can be
approximated by a finite--dimensional linear space
$L_{\varepsilon}$ with any given accuracy $\varepsilon$.

The number of coexisting inherited units (``quasi-species") is
estimated from above as $\dim L_{\varepsilon}$. This estimate is
true both for stationary and non-stationary coexistence. In its
general form this estimate was proved in 1980 \cite{GorMik,G1},
but the reasoning of this type has a long history. Perhaps,
G.~Gause \cite{Gause} was the first to suggest the direct
connection between the number of species and the number of
resources. One can call this number ``dimension of the
environment." He proposed the famous concurrent exclusion
principle. This principle is often named as the Gause principle,
but G.~Gause considered his work as development of Ch.~Darwin
ideas of the struggle for existence. This is obvious even from the
title of his book \cite{Gause}. More details about early history
of the concurrent exclusion principle are presented in the review
paper of G.~Hardin \cite{hardin}.

MacArthur and Levins \cite{MacArtLevins1964} suggested that the
number of coexisting species is limited by the number of
ecological resources. Later \cite{MacArtLevins1967}, they studied
the continuous resource distribution (niche space) where  the
number of species is limited by the fact that the niches must not
overlap too much. In 1999, G.~Meszena and J.A.J.~Metz \cite{MeMe}
developed further the idea of environmental feedback
dimensionality (perhaps, independently of \cite{GorMik,G1}). In
2003 \cite{OdoMetz}, the theory of structurally stable stationary
coexistence was developed, and in 2006 the idea of robustness in
concurrent exclusion was approached again, as a ``unified theory"
of ``competitive exclusion and limiting similarity"
\cite{MeGyPaMe}. All these achievements are related to estimation
of dimension of the set $\{k(\mu)\, | \, \mu \in M \}$ or of some
its subsets. This dimension plays the role of ``robust dimension
of population regulation".

Section~\ref{almost} contains auxiliary results from functional
analysis. Two problems are studied: (i) how to describe the sets
of global maximum points in a typical compact set of continuous
functions on a compact metric space; and (ii) how to define the
``typicalness" in Banach spaces in order to meet intuitive
expectations. It is obvious that typicalness in the sense of Baire
category violates some of the essential requirements of common
sense, for example even a real line can be divided into a set of
the zero measure and a set of first category. Hence, a stronger
notion is needed. {\it Completely thin sets} are introduced and
the typical properties of compact sets of continuous functions are
studied (the sets of exclusions are completely thin).

A theorem of selection efficiency is presented in
Section~\ref{selteor}. The sense of this theorem is as follows:
for almost every system the support of all $\omega$-limit
distributions is small (in an appropriate strong sense). Its
geometrical interpretation suggested by M. Gromov is explained in
Section~\ref{Grom}. Specific entropy--like functions, the
decreasing measures of diversity, are constructed in
Section~\ref{entropy}. Decreasing of these functions describes
self-organization.

The drift equations (Section~\ref{drift}) describe the asymptotic
behavior of systems with inheritance near the limit distributions
(when $X$ is a domain in $R^n$, or a manifold). That asymptotics
proves to be the motion of narrow, almost Gaussian peaks. The
drift equations are ODEs. In fact, the drift equations substitute
the initial infinite-dimensional dynamic system (\ref{sel1}) in
many applications: usually the system has enough time to reach the
drift asymptotic. The bifurcations with change of number of peaks
deserve special attention: this ''branching-type" evolution
\cite{Doeb}, can be related to speciation.

The simplest model for ``reproduction + small mutations" is
developed. The limit of zero mutations is singular, because
arbitrary small (but non-zero) mutations added to equation
(\ref{sel1}) destroy dynamical invariance of subspaces $\{\mu \, |
\, {\rm supp}\mu \subseteq Y\}$ for closed subsets $Y \subset X$.
Nevertheless, if we consider initial distributions $\mu_0$ with
${\rm supp}\mu=X$ (all variability is actually present), then
sufficiently small mutations change nearly nothing, just the limit
$\delta$-shaped peaks transform into sufficiently narrow peaks,
and zero limit of the velocity of their drift at $t \rightarrow
\infty$ substitutes by a small finite one in the presence of drift
effect. Moreover, there exists a {\it scale invariance}, and
dynamics for large $t$ does not depend on nonzero  mutation
intensity, if the last is sufficiently small: to change this
intensity, we need just to rescale time.

Various types of distribution stability  are studied in
Section~\ref{stab}: internal stability (stability with respect to
perturbations that do not extend the support), external stability or
uninvadability (stability to the strongly small perturbation which
extend the support), and stable realizability (stability with
respect to a small shift and small extension of density peaks). The
general condition for stable realizability is the usual ODE Lyapunov
stability condition with respect to the corresponding drift
equations.

The cell population structured by age (and age-defined variables,
size, chemical properties, etc.) is studied in
Section~\ref{cellSinch}. The most restrictive assumption in the
model is that all cells have the same cell-cycle period, $T$. Hence,
the cell-cycle phase is the inherited variable. Nevertheless, the
general results from previous sections cannot be applied to this
model because of additional symmetry. Direct analysis of the example
shows that in this case selection is also efficient and the
equivariant selection theory is possible. This selection is an
example of self-synchronization. Small deviations from the basic
assumption lead to smooth self-synchronization waves, and large
deviations can destroy the effect.

In Section~\ref{main} a brief description of the main results is
presented.

\section{Selection Theorem}

\subsection{Optimality principle for limit diversity}\label{optima}

Description of the limit behavior of a dynamical system can be
much more complicated than enumerating stable fixed points and
limit cycles. The leading rival to adequately formalize the limit
behavior is the concept of the ``$\omega$-limit set". It was
discussed in detail in the classical monograph \cite{Birkhoff}.
The fundamental textbook on dynamical systems \cite{Katok} and the
introductory review \cite{Katok2} are also available.

Let $f(t)$ be the dependence of the position of  point in the phase
space on time $t$ (i.e. the {\it motion} of the dynamical system). A
point $y$ is a $\omega$-limit point of the motion $f(t)$, if there
exists such a sequence of times $t_i \rightarrow\infty$, that
$f(t_i) \rightarrow y$.

The set of all $\omega$-limit points for the given motion $f(t)$ is
called the $\omega$-limit set. If, for example, $f(t)$ tends to the
equilibrium point $y^*$ then the corresponding $\omega$-limit set
consists of this equilibrium point. If $f(t)$ is winding onto a
closed trajectory (the limit cycle), then the corresponding
$\omega$-limit set consists of the points of the cycle and so on.

General $\omega$-limit sets are not encountered oft in specific
situations. This is because of the lack of efficient methods to
find them in a general situation. Systems with inheritance is a
case, where there are efficient methods to estimate the limit sets
from above. This is done by the optimality principle.

Let $\mu(t)$ be a solution of (\ref{sel1}). Note that
\begin{equation} \label{sel4}
\mu(t)=\mu(0) \exp \int_0^t k_{\mu (\tau)} \,  \D
 \tau.
\end{equation}
Here and below we do not display the dependence of distributions
$\mu$ and of the reproduction coefficients $k$ on $x$ when it is
not necessary. Fix the notation for the average value of $ k_{\mu
(\tau)}$ on the segment $[0,t]$
\begin{equation}\label{averages}
\langle k_{\mu(t)}\rangle _t={1 \over t} \int_0^t k_{\mu (\tau)}
\,  \D
 \tau.
\end{equation}

Then the expression (\ref{sel4}) can be rewritten as
\begin{equation} \label{sel4'}
\mu(t)=\mu(0) \exp(t\langle k_{\mu(t)}\rangle _t).
\end{equation}

If $\mu^*$ is the $\omega$-limit point of the solution $\mu(t)$,
then there exists such a sequence of times $t_i\rightarrow\infty$,
that $\mu(t_i) \rightarrow \mu^*$. Let it be possible to chose a
convergent subsequence of the sequence  of the average
reproduction coefficients $\langle k_{\mu(t)}\rangle _{t}$, which
corresponds to times $t_i$. We denote as $k^*$ the limit of this
subsequence. Then, the following statement is valid: on the
support of $\mu^*$ the function $k^*$ vanishes and on the support
of $\mu(0)$ it is non-positive:
\begin{eqnarray} \label{optsel}
k^*(x)&=&0 \; \mbox{if} \; x \in {\rm supp}\mu^*, \nonumber \\
k^*(x)&\leq& 0 \; \mbox{if} \; x \in {\rm supp}\mu(0).
\end{eqnarray}

Taking into account the fact that ${\rm supp}\mu^* \subseteq {\rm
supp}\mu(0)$, we come to the formulation of {\bf the optimality
principle} (\ref{optsel}):  {\it The support of limit distribution
consists of points of the global maximum of the average
reproduction coefficient on the initial distribution support. The
corresponding maximum value is zero. }

We should also note that not necessarily all points of maximum of
$k^*$ on ${\rm supp}\mu(0)$ belong to  ${\rm supp}\mu^* $, but all
points of ${\rm supp}\mu^*$ are the points of maximum of $k^*$ on
${\rm supp}\mu(0)$.

If $\mu(t)$ tends to the fixed point $\mu^*$, then $\langle
k_{\mu(t)}\rangle _t \rightarrow k_{\mu^*}$ as
$t\rightarrow\infty$, and ${\rm supp}\mu^*$ consists of the points
of the global maximum of the corresponding reproduction
coefficient $k_{\mu^*}$ on the support of $\mu^*$. The
corresponding maximum value is zero.

If $\mu(t)$ tends to the limit cycle $\mu^*(t)$
($\mu^*(t+T)=\mu^*(t)$), then all the distributions $\mu^*(t)$
have the same support. The points of this support are the points
of maximum (global, zero) of the averaged over the cycle
reproduction coefficient
\begin{equation}
k^*=\langle k_{\mu^*(t)}\rangle_T={1 \over T} \int_0^T k_{\mu^*
(\tau)} \,  \D \tau,
\end{equation}
on the support of $\mu(0)$.

The supports of the $\omega$-limit distributions are specified by
the functions $k^*$. It is obvious where to get these functions
from for the cases of fixed points and limit cycles. There are at
least two questions: what ensures the existence of average
reproduction coefficients at $t\rightarrow \infty$, and how to use
the described extremal principle (and how efficient is it). The
latter question is the subject to be considered in the following
sections. In the situation to follow the answers to these
questions have the validity of theorems.

Due to the theorem about weak$^*$ compactness, the set of
reproduction coefficients $k_M = \{k_{\mu}\, | \,\mu \in M \}$ is
precompact, hence, the set of averages (\ref{averages}) is
precompact too, because it is the subset of the closed convex hull
$\overline{\rm conv}(k_M)$ of the compact set. This compactness
allows us to claim the existence of the {\it average reproduction
coefficient} $k^*$ for the description of the $\omega$-limit
distribution $\mu^*$ with the optimality principle (\ref{optsel}).

\subsection{How many points \\ does the limit distribution support
hold?}\label{estimates}

The limit distribution is concentrated in the points of (zero)
global maximum of the average reproduction coefficient. The
average is taken along the solution, but the solution is not known
beforehand. With the convergence towards a fixed point or to a
limit cycle this difficulty can be circumvented. In the general
case the extremal principle can be used without knowing the
solution, in the following way \cite{G1}. Considered is a set of
all dependencies $\mu(t)$ where $\mu$ belongs to the phase space,
the bounded set $M$. The set of all averages over $t$ is
$\{\langle k_{\mu(t)}\rangle _t\}$. Further, taken are all limits
of  sequences formed by these averages -- the set of averages is
closed. The result is the closed convex hull $\overline{\rm
conv}(k_M)$ of the compact set $k_M$. This set involves all
possible averages (\ref{averages}) and all their limits. In order
to construct it, the true solution $\mu(t)$ is not needed.

{\it The weak optimality principle} is expressed as follows. Let
$\mu(t)$ be a solution of (\ref{sel1})  in $M$, $\mu^*$ is any of
its $\omega$-limit distributions. Then in the set $\overline{\rm
conv}(k_M)$ there is such a function $k^*$ that its maximum value on
the support ${\rm supp} \mu_0 $ of the initial distribution $\mu_0$
equals to zero, and ${\rm supp} \mu^*$ consists of the points of the
global maximum of $k^*$ on ${\rm supp} \mu_0$ only (\ref{optsel}).

Of course, in the set $\overline{\rm conv}(k_M)$  there are
usually many functions that are irrelevant to the time average
reproduction coefficients for the given motion $\mu(t)$.
Therefore, the weak extremal principle is really weak -- it gives
too many possible supports of $\mu^*$. However, even such a
principle can help to obtain useful estimates of the number of
points in the supports of $\omega$-limit distributions.

It is not difficult to suggest systems of the form (\ref{sel1}),
in which any set can be the limit distribution support. The
simplest example: $k_{\mu} \equiv 0$. Here $\omega$-limit (fixed)
is any distribution. However, almost any arbitrary small
perturbation of the system destroys  this pathological property.

In the realistic systems, especially in biology, the coefficients
fluctuate and are never known exactly. Moreover, the models are in
advance known to have a finite error which cannot be exterminated
by the choice of the parameters values. This gives rise to an idea
to consider not individual systems (\ref{sel1}), but ensembles of
similar systems \cite{G1}.

Let us estimate the maximum for each individual system from the
ensemble (in its $\omega$-limit distributions), and then, estimate
the minimum of these maxima over the whole ensemble -- ({\it the
minimax estimation}). The latter is motivated by the fact, that if
the inherited unit has gone extinct under some conditions, it will
not appear even under the change of conditions.

Let us consider an ensemble that is simply the
$\varepsilon$-neighborhood of the given system (\ref{sel1}). The
minimax estimates of the number of points in the support of
$\omega$-limit distribution are constructed by approximating the
dependencies $k_{\mu}$ by finite sums
\begin{equation}\label{sel7}
k_{\mu}=\varphi_0(x)+\sum_{i=1}^n \varphi_i(x) \psi_i (\mu).
\end{equation}
Here $\varphi_i$ depend on $x$ only,  and $\psi_i$ depend on $\mu$
only. Let $\varepsilon_n > 0$ be the distance from $k_{\mu}$ to
the nearest sum (\ref{sel7}) (the ``distance" is understood in the
suitable rigorous sense, which depends on the specific problem).
So, we reduced  the problem to the estimation of the diameters
$\varepsilon_n > 0$ of the set $\overline{\rm conv}(k_M)$.

{\bf The minimax estimation of the number of points in the limit
distribution support} gives the answer to the question,  ``How
many points does the limit distribution support hold": {\it If
$\varepsilon > \varepsilon_n$ then, in the $\varepsilon$-vicinity
of $k_{\mu}$, the minimum of the maxima of the number of points in
the $\omega$-limit distribution support does not exceed $n$.}

In order to understand this estimate it is sufficient to consider
system (\ref{sel1}) with $k_{\mu}$ of the form (\ref{sel7}). In this
case for any dependence $\mu(t)$ the averages (\ref{averages})  have
the form
\begin{equation}\label{sel8}
\langle k_{\mu(t)}\rangle _t={1 \over t} \int_0^t k_{\mu (\tau)}
\,  \D
 \tau =
\varphi_0(x)+\sum_{i=1}^n \varphi_i(x) a_i.
\end{equation}
where $a_i$ are some numbers. The ensemble of the functions
(\ref{sel8}) for various  $a_i$ forms a $n$-dimensional linear
manifold. How many points of the global maximum (equal to zero)
could a function of this family have?

Generally speaking, it can have any number of maxima. However, it
seems obvious, that ``usually" one function has only one point of
global maximum, while it is ``improbable" that the maximum value is
zero. At least, with an arbitrary small perturbation of the given
function, we can achieve for the point of the global maximum to be
unique and the maximum value be non-zero.

In a one-parametric family of functions there may occur zero value
of the global maximum, which cannot be eliminated by a small
perturbation, and individual functions of the family may have two
global maxima.

In the general case we can state, that ``usually" each function of
the $n$-parametric family (\ref{sel8}) can have not more than $n$
points of the zero global maximum (of course, there may be less,
and the global maximum is, as a rule, not equal to zero at all for
the majority of functions of the family). What ``usually" means
here requires a special explanation given in the next section.

In application $k_{\mu}$ is often represented by an integral
operator, linear or nonlinear. In this case the form (\ref{sel7})
corresponds to the kernels of integral operators, represented in a
form of the sums of functions' products. For example, the
reproduction coefficient of the following form
\begin{eqnarray}\label{sel9}
k_{\mu}=\varphi_0(x)+\int K(x,y) \mu(y) \,  \D y, \nonumber \\
\mbox{where} \; K(x,y)=\sum_{i=1}^n \varphi_i(x) g_i(y),
\end{eqnarray}
has also the form (\ref{sel7}) with $\psi_i (\mu)=\int g_i(y)
\mu(y) \,  \D y$.

The linear reproduction coefficients occur in applications rather
frequently. For them the problem of the minimax estimation of the
number of points in the $\omega$-limit distribution support is
reduced to the question of the accuracy of approximation of the
linear integral operator by the sums of kernels-products
(\ref{sel9}).

\subsection{Almost finite sets and ``almost always"}\label{almost}

In this section, some auxiliary propositions and definitions are
presented. The supports of the $\omega$-limit distributions for
the systems with inheritance were characterized by the optimality
principle. These supports  consist of points of global maximum of
the average reproduction coefficient. We can a priori (without
studying the solutions in details) characterize the compact set
that includes all possible average reproduction coefficients.
Hence, we get a problem: how to describe the set of global maximum
for all functions from generic compact set of functions. First of
all, any closed subset $M \subset X$ is a set of global maximum of
a continuous function, for example, of the function $f(x)=-
\rho(x,M)$, where $\rho(x,M)$ is the distance between a set and a
point: $\rho(x,M) = \inf _{y \in M} \rho(x,y)$, and $\rho(x,y)$ is
the distance between points. Nevertheless, we can expect that one
generic function has one point of global maximum, in a generic
one-parametric family might exist functions with two points of
global maximum, etc. How these expectations meet the exact
results? What does the notion ``generic" mean? What can we say
about sets of global maximum of functions from a generic compact
family? In this section we answer these questions.

``Almost always", ``typically", ``generically" a function has only
one point of global maximum. This sentence should be given an
rigorous meaning. Formally it is not difficult, but haste is
dangerous when defining ``genericity".

Here are some examples of correct but useless statements about
``generic" properties of function: Almost every continuous
function is not differentiable; Almost every $C^1$-function is not
convex. Their meaning for applications is most probably this: the
genericity used above for continuous functions or for
$C^1$-function is irrelevant to the subject.

Most frequently the motivation for definitions of genericity is
found in such a situation: given $n$ equations with $m$ unknowns,
what can we say about the solutions? The answer is: in a typical
situation, if there are more equations, than the unknowns ($n>m$),
there are no solutions at all, but if $n\leq m$ (n is less or
equal to m), then, either there is a ($m-n$)-parametric family of
solutions, or there are no solutions.

The best known example of using this reasoning is the {\it Gibbs
phase rule} in classical chemical thermodynamics. It limits the
number of co-existing phases. There exists a well-known example of
such reasoning in mathematical biophysics too. Let us consider a
medium where $n$ species coexist. The medium is assumed to be
described by $m$ parameters $s_j$. Dynamics of these parameters
depends on the organisms. In the simplest case, the medium is a
well-mixed solution of $m$ substances. Let the organisms interact
through the medium, changing its parameters -- concentrations of
$m$ substances. It can be formalized by a system of equation:
\begin{eqnarray} \label{selfinMed}
&&{\D \mu_i \over \D t}= k_i(s_1, \ldots, s_m) \times \mu_i \,
(i=1, \ldots n); \nonumber \\ &&{\D s_j \over \D t} = q_j(s_1,
\ldots, s_m, \mu_1, \ldots, \mu_n ) \, (j=1, \ldots m),
\end{eqnarray}
In a steady state, for each of the coexisting species we have an
equation with respect to the  state of the medium: the
corresponding reproduction coefficient $k_i$ is zero. So, the
number of such species cannot exceed the number of parameters of
the medium. In a typical situation, in the $m$-parametric medium
in a steady state there can exist not more than $m$ species. This
is the {concurrent exclusion principle} in the G.~Gause form
\cite{Gause}. Here, the main  hypothesis  about interaction of
organisms with the media is that the number of essential
components of the media is bounded from above by $m$ and increase
of the number of species does not extend the list of components
further. Dynamics of parameters depends on the organisms, but
their nomenclature is fixed.

This concurrent exclusion principle allows numerous
generalizations
\cite{MacArtLevins1964,MacArtLevins1967,Levin1970,OdoMetz,MeGyPaMe}.
Theorem of the natural selection efficiency may be also considered
as its generalization.

Analogous assertion for a non-steady state coexistence of species
in the case of equations (\ref{selfinMed}) is not true. It is not
difficult to give an example of stable coexistence under
oscillating conditions of $n$ species in the $m$-parametric medium
at $n>m$.

But, if $k_i(s_1, \ldots, s_m)$ are linear functions of $s_1,
\ldots, s_m$, then for non-stable conditions we have the
concurrent exclusion principle, too. In that case, the average in
time of the reproduction coefficient  is the reproduction
coefficient for the average state of the medium: $$\langle
k_i(s_1(t), \ldots, s_m(t) ) \rangle = k_i(\langle s_1 \rangle,
\ldots, \langle s_m \rangle)$$ because of linearity. If $\langle
x_i \rangle \neq 0$ then $k_i(\langle s_1 \rangle, \ldots, \langle
s_m \rangle) = 0$, and we obtain the non-stationary concurrent
exclusion principle ``in average". And again, it is valid ``almost
always".

The non-stationary  concurrent exclusion principle ``in average"
is valid for linear reproduction coefficients. This is a
combination of the  Volterra \cite{Volterra} averaging principle
and the Gause principle,

It is worth to mention that, for our basic system (\ref{sel1}), if
$k_{\mu}$ are linear functions of $\mu$, then the average in time
of the reproduction coefficient $k_{\mu(t)}$ is the reproduction
coefficient for the average $\mu(t)$ because of linearity.
Therefore, the optimality principle (\ref{optsel}) for the average
reproduction coefficient $k^*$, transforms into the following
optimality principle for the reproduction coefficient
$k_{\langle\mu\rangle}$ of the average distribution
$\langle\mu\rangle$
\begin{eqnarray}\label{averageoptim}
k_{\langle\mu\rangle}(x)&=&0 \; \mbox{if} \; x \in {\rm
supp}\mu^*, \nonumber \\ k_{\langle\mu\rangle}(x)&\leq& 0 \;
\mbox{if} \; x \in {\rm supp}\mu(0).
\end{eqnarray}
(the generalized {\it Volterra averaging principle}
\cite{Volterra}).

Formally, various definitions of genericity are constructed as
follows. All systems (or cases, or situations and so on) under
consideration are somehow parameterized -- by sets of vectors,
functions, matrices etc. Thus, the ``space of systems" $Q$ can be
described. Then the {\it  ``thin sets"} are introduced into $Q$,
i.e. the sets, which we shall later neglect. The union of a finite
or countable number of thin sets, as well as the intersection of
any number of them should be thin again, while the whole $Q$ is
not thin. There are two traditional ways to determine thinness.

\begin{enumerate}
\item{A set is considered thin when it has {\it measure zero}. This is
resonable for a finite-dimensional case, when there is the
standard Lebesgue measure -- the length, the area, the volume.}
\item{But most frequently we deal with the functional parameters.
In that case it is common to restore to the second definition,
according to which the sets of Baire first category are negligible.
The construction begins with nowhere dense sets. The set $Y$ is
nowhere dense in $Q$, if in any nonempty open set $V\subset Q$ (for
example, in a ball) there exists a nonempty open subset $W\subset V$
(for example, a ball), which does not intersect with $Y$: $W\cap Y
=\emptyset$. Roughly speaking, Y is ``full of holes" -- in any
neighborhood of any point of the set $Y$ there is an open hole.
Countable union of nowhere dense sets is called the set of first
category. The second usual way is to define thin sets as the {\it
sets of first category}. A  {\it residual set} (a ``thick" set)  is
the complement of a set of the first category.}
\end{enumerate}

For the second approach, the Baire category theorem is important: In
a non-empty complete metric space, any countable intersection of
dense, open subsets is non-empty.

But even the real line  $R$ can be divided into two sets, one of
which has zero measure, the other is of first category. The
genericity in the sense of measure and the genericity in the sense
of category considerably differ in the applications where both of
these concepts can be used. The conflict between the two main views
on genericity stimulated efforts to invent new and stronger
approaches.

Systems (\ref{sel1}) were parameterized by continuous maps $\mu
\mapsto k_{\mu} $. Denote by $Q$ the space of these maps $M
\rightarrow C(X)$ with the topology of uniform convergence on $M$.
It is a Banach space. Therefore, we shall consider below thin sets
in a Banach space $Q$. First of all, let us consider
$n$-dimensional affine compact subsets of $Q$ as a Banach space of
affine maps $\Psi: [0,1]^n \rightarrow Q$ ($\Psi(\alpha_1, \ldots
\alpha_n)=\sum_i \alpha_i f_i + \varphi$, $\alpha_i \in [0,1]$,
$f_i, \varphi \in Q$) in the maximum norm. For the image of a map
$\Psi$ we use the standard notation ${\rm im}\Psi$.
\begin{definition}
A set $Y \subset Q$ is $n$-thin, if the set of affine maps $\Psi:
[0,1]^n \rightarrow Q$ with non-empty intersection ${\rm im} \Psi
\cap Y \neq \emptyset$ is the set of first category.
\end{definition}
All compact sets in infinite-dimensional spaces and closed linear
subspaces with codimension greater then $n$ are $n$-thin. If $\dim Q
\leq n$, then only empty set is $n$-thin in $Q$. The union of a
finite or countable number of $n$-thin sets, as well as the
intersection of any number of them is $n$-thin, while the whole $Q$
is not $n$-thin.

Let us consider compact subsets in $Q$ parametrized by points of a
compact space $K$. It can be presented as a Banach space $C(K,Q)$ of
continuous maps $K \rightarrow Q$ in the maximum norm.
\begin{definition}\label{complthin}
A set $Y \subset Q$ is completely thin, if for any compact $K$ the
set of continuous maps $\Psi: K \rightarrow Q$ with non-empty
intersection ${\rm im} \Psi \cap Y \neq \emptyset$ is the set of
first category.
\end{definition}

A set $Y$ in the Banach space $Q$ is completely thin, if for any
compact set $K$ in $Q$ and arbitrary positive $\varepsilon>0$ there
exists a vector $q \in Q$, such that $\|q\|<\varepsilon$ and $K+q$
does not intersect $Y$: $(K+q) \cap Y =\emptyset$. All compact sets
in infinite-dimensional spaces and closed linear subspaces with
infinite codimension  are completely thin. Only empty set is
completely thin in a finite-dimensional space. The union of a finite
or countable number of completely thin sets, as well as the
intersection of any number of them is completely thin, while the
whole $Q$ is not completely thin.

\begin{proposition}\label{pro:comcom}
If a set $Y$ in the Banach space $Q$ is completely thin, then for
any compact metric space $K$ the set of continuous maps $\Psi: K
\rightarrow Q$ with non-empty intersection ${\rm im} \Psi \cap Y
\neq \emptyset$ is completely thin in the Banach space $C(K,Q)$.
\end{proposition}

To prove this proposition it is sufficient to mention that for any
compact $P$ the space $C(P,C(K,Q))$ of continuous maps $P
\rightarrow C(K,Q)$ is isomorphic to the space $C(P\times K,Q)$, and
$P\times K$ is compact again.

Below the wording ``almost always" means: the set of exclusions is
completely thin. The main result presented in this section sounds as
follows: almost always the sets of global maxima of  functions from
a compact set are uniformly almost finite.

\begin{proposition}\label{pro:notdense}
Let $X$ have no isolated points. Then almost always a function $f
\in C(X)$ has nowhere dense set of zeros $\{x \in X \, | \,
f(x)=0 \}$ (the set of exclusions is completely thin in $C(X)$).
\end{proposition}
In order to prove this proposition, let us mention that the
topology of $X$ has a countable base $\{U_i\}_{i=1}^\infty$. For
any $i$ the set of functions $${\rm Annul}U_i=\{f \in C(X) \, | \,
f(x) = 0 \; \mbox{for all} \; x \in U_i \}$$ is closed subspace of
$C(X)$ (even an ideal) with infinite codimension. For any compact
set $K \subset C(X)$ the sets of shift vectors $$V_i =\{y \in
C(X)\, | \,(K+y)\cap {\rm Annul}U_i = \emptyset\}$$ is open and
dense in $C(X)$. Hence, the set of shifts $\cap_i V_i$ is dense
residual set in $C(X)$.

After combination Proposition~\ref{pro:notdense} with
Proposition~\ref{pro:comcom} we get the following
\begin{proposition}\label{pro:notdenseset}
Let $X$ have no isolated points. Then for any compact space $K$
and almost every continuous map $\Psi: K \rightarrow C(X)$ all
functions $f \in {\rm im } \Psi$ have nowhere dense sets of zeros
(the set of exclusions is completely thin in $C(K,C(X))$).
\end{proposition}
In other words, in almost every compact family of continuous
functions all the functions have nowhere dense sets of zeros.

In construction of ``almost finite" sets we follow a rather old
idea that was used by Liouville in construction of his ``almost
rational" transcendental numbers \cite{Liouville}. A Liouville
number is a transcendental number which has very close rational
number approximations. An irrational number $\beta$ is called a
Liouville number if, for each $n$, there exist integers $p$ and
$q$ such that

$$\left|\beta - \frac{p}{q} \right| < \frac{1}{q^n}$$

An example of such a number gives Liouville's constant, sometimes
also called Liouville's number, that is the real number defined by

$$L = \sum_{n=1}^{\infty} 10^{-n!}$$

Liouville's constant is a decimal fraction with a 1 in each
decimal place corresponding to a factorial, and zeros everywhere
else. It was the first decimal constant to be proven
transcendental.

In some sense, almost all real numbers are the Liouville numbers:
the set of all Liouville numbers is the residual set. To prove this
statement let us enumerate all the rational numbers: $r_1,\, r_2,\,
\ldots$, $r_n =p_n/q_n$. The following set $U_{\epsilon}$ is open
and dense on the real line:
\begin{equation}
U_{\epsilon}= \bigcup_{n=1}^{\infty} \left\{\beta \, | \,
\left|\beta - \frac{p_n}{q_n}\right| <
\frac{\epsilon}{q_n^n}\right\}.
\end{equation}
The intersection of $U_{\epsilon}$ for $\epsilon = 1/2,\, 1/4, \,
1/8, \, \ldots$
\begin{equation}
U= \bigcap_{k=1}^{\infty} U_{\frac{1}{2^k}}
\end{equation}
is the residual set, and numbers from $U$ are the Liouville numbers.
On the other hand, $U$ has zero Lebesgue measure, and it gives us an
example of dividing the real line on the set of first category (the
complement of $U$) and the set of the zero measure $U$.

Let us consider a space of closed subsets of the compact metric
space $X$ endowed by the Hausdorff distance. The Hausdorff
distance between closed subsets of $X$ is
\begin{equation}
{\rm dist}(A,B)=\max\{\sup_{x \in A} \inf_{x \in B} \rho(x,y) ,
\sup_{x \in B} \inf_{x \in A} \rho(x,y) \}.
\end{equation}

The almost finite sets were introduced in \cite{G1} for
description of the typical sets of maxima for continuous functions
from a compact set. This definition depends on an arbitrary
sequence $\varepsilon_n > 0$, $\varepsilon_n \rightarrow 0$. For
any such sequence we construct a class of subsets $Y\subset X$
that can be approximated by finite set faster than $\varepsilon_n
\rightarrow 0$, and for families of sets we introduce a notion of
{\it uniform} approximation by finite sets faster than
$\varepsilon_n \rightarrow 0$:

\begin{definition}\label{almfin}
Let $\varepsilon_n > 0$, $\varepsilon_n \rightarrow 0$. The set
$Y\subset X$ can be approximated by finite sets faster than
$\varepsilon_n \rightarrow 0$ ($\varepsilon_n > 0$), if for any
$\delta > 0$ there exists a finite set $S_N$ such that ${\rm
dist}(S_N, Y)<\delta \varepsilon_N$. The  sets of family
$\mathbb{Y}$ can be uniformly approximated by finite sets faster
than $\varepsilon_n \rightarrow 0$, if for any $\delta > 0$ there
exists such a number $N$ that for any $Y \in \mathbb{Y}$ there
exists a finite set $S_N$ such that ${\rm dist}(S_N, Y)<\delta
\varepsilon_N$.
\end{definition}

The simplest example of almost finite set on the real line for a
given $\varepsilon_n \rightarrow 0$ ($\varepsilon_n > 0$) is the
sequence $\varepsilon_n / n $. If $\varepsilon_n < {\rm const}/n$,
then the set $Y$ on the real line which can be approximated by
finite sets faster than $\varepsilon_n \rightarrow 0$ have zero
Lebesgue measure. At the same time, it is nowhere dense, because it
can be covered by a finite number of intervals with an arbitrary
small sum of lengths (hence, in any interval we can find a
subinterval free of points of $Y$).

Let us study the sets of global maxima ${\rm argmax} f$ for
continuous functions $f \in C(X)$. For each $f \in C(X)$ and any
$\epsilon>0$ there exists $\phi \in C(X)$ such that $\|f-\phi \|
\leq \epsilon$ and ${\rm argmax} \phi$ consists of one point. Such
a function  $\phi$ can be chosen in the form
\begin{equation}
\phi (x) = f(x) + \frac{\epsilon}{1+\rho(x,x_0)^2},
\end{equation}
where $x_0$ is an arbitrary element of ${\rm argmax} f$. In this
case ${\rm argmax} \phi = \{x_0\}$.

Hence, the set ${\rm argmax} f$ can be reduced to one point by an
arbitrary small perturbations of the function $f$. On the other
hand, it is impossible to extend significantly the set ${\rm
argmax} f$ by a sufficiently small perturbation, the dependence of
this set on $f$ is semicontinuous in the following sense.

\begin{proposition}\label{pro:semicon}
For given $f \in C(X)$ and any $\varepsilon > 0$ there exists
$\delta
> 0$ such that, whenever $\|f - \phi\| < \delta$, then
\begin{equation}\label{semicont}
\max_{x \in {\rm argmax} \phi} \min_{y \in {\rm argmax}  f}
\rho(x,y) < \varepsilon.
\end{equation}
\end{proposition}

In order to derive the dependence of $\delta$ on $\varepsilon$ in
Proposition~\ref{pro:semicon}, we can use the following auxiliary
function:
\begin{equation}
\eta_f(r) = \min_{\rho(x,{\rm argmax} f)\geq r}\left\{\max_X f -
f(x) \right\},
\end{equation}
where $\rho(x,{\rm argmax} f)$ is the distance between a set and a
point.

The function $\eta_f(r)$ is monotone nondecreasing, $\eta_f(0)=0$,
and $\eta_f(r)>0$ for $r>0$. We can take in
Proposition~\ref{pro:semicon} any $\delta < \eta_f(\varepsilon)$,
for example, $\delta = \eta_f(\varepsilon)/2$.

In particular, if the set ${\rm argmax} f$ consists of one point
$x_0$, then for sufficiently small perturbations of $f$ the set
${\rm argmax}$ remains in an arbitrary small ball near $x_0$.

These constructions can be generalized onto $n$-parametric affine
compact families of continuous functions. Let us consider affine
maps of the cube $[0,1]^k$ into $C(X)$, $\Phi:[0,1]^k \rightarrow
C(X)$. The space of all such maps is a Banach space endowed with
the maximum norm.

\begin{proposition}\label{pro:netred}
For any affine map $\Phi :[0,1]^k \rightarrow C(X)$ and an
arbitrary $\epsilon > 0$ there exists such a continuous function
$\psi \in C(X)$, that $\|\psi\| < \epsilon$ and the set ${\rm
argmax}(f+\psi)$ includes not more than $k+1$ points for all $f
\in {\rm im}\Phi$.
\end{proposition}

To prove this Proposition~\ref{pro:netred} we can use the
following Lemma.
\begin{lemma}\label{lem:netred}
Let $Q \subset C(X)$ be a compact set of functions, $\varepsilon
>0$. Then there are a finite set $Y \subset X$ and a function
$\phi \in C(X)$ such that $\| \phi \| < \varepsilon $, and any
function $f \in Q + \phi$ achieves its maximum only on $Y$: ${\rm
argmax} f \subset Y$.
\end{lemma}

To find the shift function $\phi$ let us use the following
auxiliary functions: for $f \in C(X)$
\begin{equation}
\gamma _f(r) = \max_{\rho(x,y)\leq r}|f(x)-f(y)|.
\end{equation}
The function $\gamma _f(r)$ is monotone nondecreasing, $\gamma
_f(0)=0$, and $\gamma _f(r)\rightarrow 0$ for $r \rightarrow 0$.
Sometimes one calls it the {\it uniform continuity module} of $f$.
For the compact of functions $Q \subset C(X)$ we can also define
the uniform continuity module:
\begin{equation}
\gamma_Q(r) = \max_{f \in Q}\gamma _f(r) = \max_{f \in Q, \,
\rho(x,y)\leq r}|f(x)-f(y)|.
\end{equation}
The function $\gamma _Q(r)$ is monotone nondecreasing, $\gamma
_Q(0)=0$, and $\gamma _Q(r)\rightarrow 0$ for $r \rightarrow 0$.
For general compact space $X$ the function $\gamma _f(r)$, $\gamma
_Q(r)$ might be not continuous at points $r \neq 0$. Instead of
them we can use their continuous majorants, for example, the
concave closures of these functions. Hypograph of a function
$\gamma(r)$, denoted ${\rm hyp}(\gamma)$, is the set $\{(r, g) \,
| \, g \leq \gamma(r)\}$. Note that the hypograph is the region
below the graph of $\gamma$. The concave closure of $\gamma(r)$
(denoted as $\overline{\rm conc} \gamma$) is the function that has
as hypograph the closure of the convex hull of ${\rm hyp}(\gamma)$
(that is, the smallest closed and convex set containing ${\rm
hyp}(\gamma)$). If a function $\gamma(r)$ on an interval $[0,R]$
is monotone nondecreasing, bounded, $\gamma (0)=0$, and $\gamma
(r)\rightarrow 0$ for $r \rightarrow 0$ (that is $\gamma(r)$ is
continuous at the point $r=0$), then the function $\overline{\rm
conc} \gamma$ on the interval $[0,R]$ is continuous, monotone
nondecreasing, $\overline{\rm conc} \gamma(r) \geq \gamma(r)$ for
all $r \in [0,R]$, $\overline{\rm conc} \gamma(0) = \gamma(0) =0$
and $\overline{\rm conc} \gamma(R) = \gamma(R)$.

For given $\gamma>0$, $\gamma < \max \gamma_Q(r)$ we can find
$r(\gamma)$, a unique solution of the equation $\overline{\rm
conc} \gamma_Q(r) = \gamma$.

For the given $\varepsilon > 0$, let us find in $X$ a finite
$r(\varepsilon /2)$-net $\{x_1,x_2, \ldots , x_N \} \subset X$.
For each $x_i$ we define a $\varepsilon$-small  ``cap function"
\begin{eqnarray}
&&\phi_i(x) = \frac{4}{3} \left(\frac{\varepsilon}{2}-
\gamma_Q\left(\rho(x,x_i)\frac{r_1}{r_2}\right)\right) \;
\mbox{if} \;\rho(x,x_i) < r_2, \nonumber \\  && \phi_i(x) = 0 \;
\mbox{if} \;\rho(x,x_i) \geq r_2,
\end{eqnarray}
where $r_1 =r(\varepsilon /2)$, $r_2=\frac{1}{2} \min \left\{r_1,
\frac{1}{2} \min_{i \neq j}\rho(x_i,x_j) \right\}$.

If $\rho(x,x_i) \leq \varepsilon /2$ and $f \in Q$, then
$|f(x)-f(x_i)|<\phi_i(x)$. If $\phi_i(x) \neq 0$ then $\phi_j(x) =
0$ for all $j \neq i$; $\|\phi_i(x)\| < \varepsilon$ and
$\phi_i(x) \geq 0$ for each $i$.

We can define the shift function $\phi$ in Lemma~\ref{lem:netred}
as

\begin{equation}
\phi(x) = \sum_{i=1}^N \phi_i(x).
\end{equation}

Lemma~\ref{lem:netred} allows us to reduce some problems
concerning global maxima of continuous functions from a compact
sets $Q \subset C(X)$ to questions about functions on finite
subsets in $X$. In particular, Proposition~\ref{pro:netred}
reduces to a question about existence of non-trivial solutions for
finite systems of linear equations. Let us consider functions on
the finite net $\{x_1,x_2, \ldots , x_N \} \subset X$. For each
affine map $\Phi :[0,1]^k \rightarrow C(X)$ ($\Phi(\alpha_1,
\ldots \alpha_n)=\sum_i \alpha_i f_i + \varphi$, $\alpha_i \in
[0,1]$, $f_i, \varphi \in C(X)$) the values $\Phi(\alpha_1, \ldots
\alpha_n)(x_i)$ are linear (non-homogeneous) functions on the
$k$-cube $[0,1]^k$. For any $q$ different points $x_{i_1}, \ldots
x_{i_q}$ of the net the system of equations
\begin{equation}
\Phi(\alpha_1, \ldots \alpha_n)(x_{i_1})=\ldots =\Phi(\alpha_1,
\ldots \alpha_n)(x_{i_q})
\end{equation}
can, generically, have a solution in the $k$-cube $[0,1]^k$ only
for $q \leq k+1$. The degenerated case with solutions for $q> k+1$
can be destroyed by an arbitrary small perturbation. This simple
remark together with Lemma~\ref{lem:netred} proves
Proposition~\ref{pro:netred}.

Note, that Proposition~\ref{pro:netred} and Lemma~\ref{lem:netred}
demonstrate us different sources of discreteness: in
Lemma~\ref{lem:netred} it is the approximation of a compact set by
a finite net, and in Proposition~\ref{pro:netred} it is the
connection between the number of parameters and the possible
number of global maximums in a $k$-parametric family of functions.
There is no direct connection between $N$ and $k$ values, and it
might be that $N\gg k$. For smooth functions in finite-dimensional
real space polynomial approximations can be used instead of
Lemma~\ref{lem:netred} in order to prove the analogue of
Proposition~\ref{pro:netred}.

For any compact $K$ the space of continuous maps $C(K,C(X))$ is
isomorphic to the space of continuous functions $C(K \times X)$.
Each continuous map $F:K \rightarrow C(X)$ can be approximated
with an arbitrary accuracy $\varepsilon
> 0$ by  finite sums of the following form ($k \geq 0$):
\begin{eqnarray}\label{finsum}
&&F(y)(x)=\sum_{i=1}^k \alpha_i(y)f_i(x) + \varphi(x) + o,
\nonumber
\\ && y\in K, \, x\in X, \, 0\leq \alpha_i \leq 1, \, f_i,\varphi \in
C(X), \, |o|< \varepsilon.
\end{eqnarray}
Each set $f_i,\varphi \in C(X)$ generates a map $\Phi :[0,1]^k
\rightarrow C(X)$. A dense subset in the space of these maps
satisfy the statement of Proposition~\ref{pro:netred}: each
function from ${\rm im}\Phi$ has not more than $k+1$ points of
global maximum. Let us use for this set of maps $\Phi$ notation
$\mathbf{P}_k$, for the correspondent set of the maps $F:K
\rightarrow C(X)$, which have the form of finite sums
(\ref{finsum}), notation $\mathbf{P}_k^K$, and $\mathbf{P}^K =
\cup_k \mathbf{P}_k^K$.

For each $\Phi \in \mathbf{P}_k^K$ and any $\varepsilon > 0$ there
is $\delta = \delta_{\Phi}(\varepsilon) > 0$ such that, whenever
$\|\Psi - \Phi\| < \delta_{\Phi}(\varepsilon)$, the set ${\rm
argmax } f$ belongs to a union of $k+1$ balls of radius
$\varepsilon$ for any $f \in {\rm im}\Psi$
(Proposition~\ref{pro:semicon}).

Let us introduce some notations: for $k \geq 0$ and $\varepsilon >
0$ $$ \mathbf{U}_{k, \varepsilon}^K= \{\Psi \in C(K,C(X))\, | \,
|\Psi - \Phi\| < \delta_{\Phi}(\varepsilon) \; \mbox{for some} \;
\Phi \in \mathbf{P}_k^K \};$$  for $\varepsilon_i >0$,
$\varepsilon_i \rightarrow 0$

$$\mathbf{V}_{ \{\varepsilon_i\}}^K= \bigcup_{k=0}^{\infty}
\mathbf{U}_{k, \varepsilon_k}^K;$$ and, finally,

$$\mathbf{W}_{ \{\varepsilon_i\}}^K =\bigcap_{s=1}^{\infty}
\mathbf{V}_{ \{\ \frac{1}{2^s} \varepsilon_i\}}^K.$$

The set $\mathbf{P}^K$ is dense in $C(K,C(X))$. Any $F \in
\mathbf{P}^K$ has the form of finite sum (\ref{finsum}), and any
$f \in {\rm im} F$ has not more than $k+1$ point of global
maximum, where $k$ is the number of summands in presentation
(\ref{finsum}). The sets $\mathbf{V}_{ \{\varepsilon_i\}}^K$ are
open and dense in the Banach space $C(K,C(X))$ for any sequence
$\varepsilon_i >0$, $\varepsilon_i \rightarrow 0$. The set
$\mathbf{W}_{ \{\varepsilon_i\}}^K $ is intersection of countable
number of open dense sets. For any $F \in \mathbf{W}_{
\{\varepsilon_i\}}^K$ the sets of the family $\{{\rm argmax}f \, |
\, f \in {\rm im} F \}$ can be uniformly approximated by finite
sets faster than $\varepsilon_n \rightarrow 0$. It is proven that
this property is typical in the Banach space $C(K,C(X))$ in the
sense of category. In order to prove that the set of exclusions is
completely thin in $C(K,C(X))$ it is sufficient to use the
approach of Proposition~\ref{pro:comcom}. Note that for arbitrary
compact space $Q$ the set of continuous maps $Q \rightarrow
C(K,C(X))$ in the maximum norm is isomorphic to the spaces $C(Q
\times K,C(X))$ and $C(Q \times K \times X)$. The space $Q \times
K$ is compact. We can apply the previous construction to the space
$C(Q \times K,C(X))$ for arbitrary compact $Q$ and get the result:
the set of exclusion is completely thin in $C(K,C(X))$.

In the definition of $\mathbf{W}_{ \{\varepsilon_i\}}^K$ we use
only one sequence $\varepsilon_i > 0$, $\varepsilon_i \rightarrow
0$. Of course, for any finite or countable set of sequences the
intersection of correspondent sets $\mathbf{W}_{
\{\varepsilon_i\}}^K$ is also a residual set, and we can claim
that almost always the sets of $\{{\rm argmax}f\, | \, f \in {\rm
im} F \}$ can be uniformly approximated by finite sets faster than
$\varepsilon_n \rightarrow 0$ for all given sequences. Let us
mention that the set of all recursive enumerable countable sets is
also countable and not continuum. This observation is very
important for algorithmic foundations of probability theory
\cite{Levin}. Let $\LL$ be a set of all sequences of real numbers
$\varepsilon_n>0$, $\varepsilon_n \rightarrow 0$ with the
property: for each $\{\varepsilon_n\} \in \LL$ the rational
hypograph $\{(n,r)\, | \,\varepsilon_n > r \in \QQ \}$ (where
$\QQ$ is the set of rational numbers)  is recursively enumerable.
The set $$\mathbf{W}_{\LL}^K =\bigcap_{\{\varepsilon_i\} \in \LL}
\mathbf{W}_{ \{\varepsilon_i\}}^K$$ is a residual set again and
almost always the sets of $\{{\rm argmax}f\, | \, f \in {\rm im} F
\}$ can be uniformly approximated by finite sets faster than
$\varepsilon_n \rightarrow 0$ for all given sequences from $\LL$.

\subsection{Selection efficiency}\label{selteor}

The first application of the extremal principle for the
$\omega$-limit sets is the theorem of the selection efficiency.
The dynamics of a system with inheritance leads indeed to a
selection in the limit $t \rightarrow \infty$. In the typical
situation, a diversity in the limit $t \rightarrow \infty$ becomes
less than the initial diversity. There is an efficient selection
for the ``best". The basic effects of selection are formulated
below. Let $X$ be compact metric space without isolated points.

\begin{theorem}{\bf(Theorem of selection efficiency.)}
\begin{enumerate}
\item{For almost every system (\ref{sel1}) the support of any $\omega$-limit distribution
is nowhere dense in $X$ (and it has the Lebesgue measure zero for
Euclidean space).}
\item{Let $\varepsilon_n>0$, $\varepsilon_n
\rightarrow 0$ be an arbitrary chosen sequence. The following
statement is true for almost every system (\ref{sel1}). Let the
support of the initial distribution be the whole $X$. Then the
support of any $\omega$-limit distribution can be approximated by
finite sets uniformly faster than $\varepsilon_n \rightarrow 0$.}
\end{enumerate}
The set of exclusive systems that do not satisfy the statement 1
or 2 is completely thin.
\end{theorem}

These properties hold for the continuous reproduction
coefficients. It is well-known, that it is dangerous to rely on
the genericity among continuous functions. For example, almost all
continuous functions are nowhere differentiable. But the
properties 1, 2 hold also for the smooth reproduction coefficients
on the manifolds and sometimes allow to replace the ``almost
finiteness" by simply finiteness.

To prove the first statement, it is sufficient to refer to
Proposition~\ref{pro:notdenseset}. In order to clarify the second
part of this theorem, note that:

\begin{enumerate}
\item{Support of an arbitrary  $\omega$-limit distribution $\mu^*$
consist of points of global maximum of the average reproduction
coefficient on a support of the initial distribution. The
corresponding maximum value is zero.} \item{Almost always a
function has only one point of global maximum, and corresponding
maximum value is not 0.} \item{In a one-parametric family of
functions almost always there may occur zero values of the global
maximum (at one point), which cannot be eliminated by a small
perturbation, and individual functions of the family may stably
have two global maximum points.} \item{For a generic $n$-parameter
family of functions, there may exist stably a function with $n$
points of global maximum and with zero value of this maximum.}
\item{Our phase space $M$ is compact. The set of corresponding
reproduction coefficients $k_M$ in $C(X)$ for the given map $\mu
\rightarrow k_{\mu}$ is compact too. The average reproduction
coefficients belong to the closed convex hull of this set
$\overline{\rm conv}(k_M)$. And it is compact too.} \item{A
compact set in a Banach space can be approximated by compacts from
finite-dimensional linear manifolds. Generically,  the function
from such a compact can have not more than $n$ points of global
maximum with zero value, where $n$ is the dimension of manifold.}
\end{enumerate}

The rest of the proof of the second statement is purely technical.
Some technical details are presented in  the previous section. The
easiest demonstration of the ``natural" character of these
properties is the demonstration of instability of exclusions: If,
for example, a function has several points of global maxima, then
with an arbitrary small perturbation (for all usually used norms)
it can be transformed into a function with the unique point of
global maximum. However ``stable" does not always mean ``dense".
The discussed properties of the system (\ref{sel1}) are valid in a
very strong sense: the set of exclusion is completely thin.

\subsection{Gromov's interpretation of selection
theorems}\label{Grom}

In his talk \cite{GroGo}, M. Gromov offered a geometric
interpretation of the selection theorems. Let us consider
dynamical systems in the standard $m$-simplex $\sigma_m$ in
$m+1$-dimensional space $R^{m+1}$: $$\sigma_m=\{x \in R^{m+1} \, |
\, x_i \geq 0, \: \sum_{i=1}^{m+1} x_i = 1 \}.$$ We assume that
simplex $\sigma_m$ is positively invariant with respect to these
dynamical systems: if the motion starts in $\sigma_m$ at some time
$t_0$, then it remains in $\sigma_m$ for $t>t_0$. Let us consider
the motions that start in the simplex $\sigma_m$ at $t=0$ and are
defined for $t>0$.

For large $m$, almost all volume of the simplex $\sigma_m$ is
concentrated in a small neighborhood of the center of $\sigma_m$,
near the point $c= \left({1 \over m}, {1 \over m}, \ldots, {1 \over
m}\right)$. Hence, one can expect that a typical motion of a general
dynamical system in $\sigma_m$ for sufficiently large $m$ spends
almost all the time in a small neighborhood of $c$.

Indeed, the $m$-dimensional volume of $\sigma_m$ is $V_m =
\frac{1}{m!}$. The part of $\sigma_m$, where $x_i \geq \varepsilon$,
has the volume $(1-\varepsilon)^mV_m$. Hence, the part of
$\sigma_m$, where $x_i < \varepsilon$ for all $i=1,\ldots , m+1$,
has the volume $V_{\varepsilon} > (1-(m+1)(1-\varepsilon)^m)V_m$.
Note, that $(m+1)(1-\varepsilon)^m \sim m\exp(-\varepsilon m)
\rightarrow 0$, if $m\rightarrow \infty$  ($1> \varepsilon > 0$).
Therefore, for $m \rightarrow \infty$, $V_{\varepsilon} =
(1-o(1))V_m$. The volume $W_{\rho}$ of the part of $\sigma_m$ with
Euclidean distance to the center $c$ less than $\rho>0$ can be
estimated as follows: $W_{\rho}
> V_{\varepsilon}$ for $\varepsilon \sqrt{m+1} = \rho$, hence
$W_{\rho} > (1-(m+1)(1-\rho/\sqrt{m+1})^m)V_m$. Finally,
$(m+1)(1-\rho/\sqrt{m+1})^m \sim m\exp(-\rho\sqrt{m})$, and
$W_{\rho}=(1-o(1))V_m$ for $m\rightarrow \infty$. Let us mention
here the opposite concentration effect for a $m$-dimensional ball
$B_m$: for $m \rightarrow \infty$ the most part of its volume is
concentrated in an arbitrary small vicinity of its boundary, the
sphere. This effect is the essence of the famous equivalence of
microcanonical and canonical ensembles in statistical physics (for
detailed discussion see \cite{Order}).

Let us consider dynamical systems with an additional property
(``inheritance"): all the faces of the simplex $\sigma_m$ are also
positively invariant with respect to the systems with inheritance.
It means that if some $x_i=0$ initially at the time $t=0$, then
$x_i=0$ for $t>0$ for all motions in $\sigma_m$. The essence of
selection theorems is as follows: a typical motion of a typical
dynamical system with inheritance spends almost all the time in a
small neighborhood  of low-dimensional faces, even if it starts near
the center of the simplex.

Let us denote by $\partial_r \sigma_m$ the union of all
$r$-dimensional faces of $\sigma_m$. Due to the selection
theorems, a typical motion of a typical dynamical system with
inheritance spends almost all time in a small neighborhood of
$\partial_r \sigma_m$ with $r\ll m$. It should not obligatory
reside near just one face from $\partial_r \sigma_m$, but can
travel in neighborhood of different faces from $\partial_r
\sigma_m$ (the drift effect). The minimax estimation of the number
of points in $\omega$-limit distributions through the diameters
$\varepsilon_n > 0$ of the set $\overline{\rm conv}(k_M)$  is the
estimation of $r$.

\subsection{Decreasing measures of diversity, \\ Lyapunov
functionals, and Burg Entropy} \label{entropy}

The distinguished Lyapunov functionals play important role in
kinetics. For physical and chemical systems such a functional is, as
a rule, the entropy or some of related functionals. The standard
examples are the free (or Helmholz) energy and the free enthalpy (or
Gibbs energy). It appears that for system (\ref{sel1}) there exist
generically a plenty of Lyapunov functionals. They can be considered
as the {\it decreasing measures of diversity}. These functionals are
very similar to the entropy, but rather to the Burg entropy
\cite{Burg,GKBurg}, and not to the classic Boltzmann--Gibbs--Shannon
entropy.

Generically, we can assume that the convex compact set
$\overline{\rm conv}(k_M)$ does not include zero. The set of
exclusion from this rule is completely thin. Then there exists a
continuous functional $l$ on $C(X)$ with positive values on
$\overline{\rm conv}(k_M)$. It is a measure on $X$ (by definition).
Of course, values of $\rho$ outside ${\rm supp} \mu(0)$ do not have
any relation to reality, and it is sufficient to discuss only the
case ${\rm supp} \mu(0)= X$. Any solution of (\ref{sel1}) can be
presented in a form: $\mu(t)=\rho(t) \mu(0)$, where $\rho(t) \in
C(X)$, $\rho(t) > 0$ at each $t$, and
\begin{equation}\label{logarifmmery}
{\D (\ln \rho(t)) \over \D t}= k_{\mu (t)}.
\end{equation}
Hence, the ``entropy"
\begin{equation}
S_l[\rho(t)]= - [l, \ln \rho(t)]
\end{equation}
monotonically decrease:
\begin{equation}\label{rostentropii}
{\D S_l[\rho(t)] \over \D t}= - [l, k_{\mu (t)}] < 0.
\end{equation}
In order to avoid the dependence of an initial distributions
$\mu(0)$ we can restrict the initial system (\ref{sel1}) onto its
invariant subspace, the space of measures which have a form $\mu=
\rho \mu^0$, where $\mu^0$ is a given measure with ${\rm supp}
\mu^0= X$, and $\rho \in C(X)$.

The space $L^2_{\mu^0}(X)$ is the completion of $C(X)$ with
respect to the norm
\begin{equation}
\| f \| =  [\mu^0, f^2]^{1/2}.
\end{equation}
It is the Hilbert space. For the scalar product we shall use the
notation $(\, , \, )_{\mu^0}$

The compact convex set $\overline{\rm conv}(k_M)$ is also compact in
$L^2_{\mu^0}(X)$. The set $$D_{\mu^0} = \left\{ \varphi \in
L^2_{\mu^0}(X) : (\varphi , f)_{\mu^0} > 0 \, \mbox{for any} \, f
\in \overline{\rm conv}(k_M) \right\}$$ is open in $L^2_{\mu^0}(X)$.
Generically, it is nonempty, and, hence, there is a dense set of
continuous functions in $D_{\mu^0}$. For any function $g \in
D_{\mu^0}$ we can define the related ``entropy"

\begin{equation}\label{entropiika}
S[\rho]=- (\ln \rho,g)_{\mu^0} = -[\mu^0, g \ln \rho].
\end{equation}
This entropy is the  average of the density logarithm with a weight
$-g$,  the set of allowed weights depends  on reproduction
coefficient. The entropy (\ref{entropiika}) decrease for each
solution of (\ref{sel1}) that has a form $\mu(t)= \rho(t) \mu^0$
with positive initial condition ($\rho(0)$ is strictly positive
function).

The introduced entropies decrease monotonically to minus infinity.
It is clear, all measures of diversity, including the classical
entropy, should  decrease in a result of the selection process. The
only question was about the monotonicity of this decreasing.

\section{Drift and mutations}

\subsection{Drift equations}\label{drift}

So far,  we talked about the support of an individual
$\omega$-limit distribution. For almost all systems it is small.
But this does not mean, that the union of these supports is small
even for one solution $\mu(t)$. It is possible that a solution is
a finite set of narrow peaks getting in time more and more narrow,
moving slower and slower, but not tending to fixed positions,
rather continuing to move along its trajectory, and the path
covered tends to infinity as $t \rightarrow \infty$.

This effect was not discovered for a long time because the slowing
down of the peaks was thought as their tendency to fixed positions
For the best of our knowledge, the first detailed publication of
the drift equations and corresponded  types of stability appeared
in book \cite{G1}, first examples of coevolution drift on a line
were published in the series of papers \cite{RozSed}.

There are other difficulties related to the typical properties of
continuous functions, which are not typical for the smooth ones.
Let us illustrate them for the distributions over a straight line
segment. Add to the reproduction coefficients $k_{\mu}$ the sum of
small and narrow peaks located on a straight line distant from
each other much more than the peak width (although it is
$\varepsilon$-small). However small is chosen the peak's height,
one can choose their width and frequency on the straight line in
such a way that from any initial distribution $\mu _0$ whose
support is the whole segment, at $t \rightarrow \infty$ we obtain
$\omega$-limit distributions, concentrated at the points of
maximum of the added peaks.

Such a model perturbation is small in the space of continuous
functions. Therefore, it can be put as follows: {\it by small
continuous perturbation the limit behavior of system (\ref{sel1})
can be reduced onto a  $\varepsilon$-net for sufficiently small
$\varepsilon$}. But this can not be done with the small smooth
perturbations (with small values of the first and the second
derivatives) in the general case. The discreteness of the net,
onto which the limit behavior is reduced by small continuous
perturbations, differs from the discreteness of the support of the
individual $\omega$-limit distribution. For an individual
distribution the number of points is estimated, roughly speaking,
by the number of essential parameters (\ref{sel7}), while for the
conjunction of limit supports -- by the number of stages in
approximation of $k_{\mu}$ by piece-wise constant functions.

Thus, in a typical case the dynamics of systems (\ref{sel1}) with
smooth reproduction coefficients transforms  a smooth initial
distributions into  the ensemble of narrow peaks. The peaks become
more narrow, their motion  slows down, but not always they tend to
fixed positions.

The equations of motion for these peaks can be obtained in the
following way \cite{G1}. Let $X$ be a domain in the
$n$-dimensional real space, and the initial distributions $\mu_0$
be assumed to have smooth density. Then, after sufficiently large
time $t$, the position of distribution peaks  are the points of
the average reproduction coefficient maximum $\langle k_{\mu}
\rangle _t$ (\ref{averages}) to any accuracy set in advance. Let
these points of maximum be $x^{\alpha}$, and
$$q_{ij}^{\alpha}= \left. - t {\partial ^2 \langle k_{\mu} \rangle _t \over \partial x_i \partial
x_j}\right|_{x=x^{\alpha}}.$$ It is easy to derive the following
differential relations
\begin{eqnarray}\label{selrela}
\sum_j q_{ij}^{\alpha} {\D x_j^{\alpha} \over \D t}& =& \left.
{\partial k_{\mu(t)} \over \partial x_i} \right|_{x=x^{\alpha}} ;
\nonumber \\ {\D q_{ij}^{\alpha} \over \D t}& = &
\left.-{\partial^2 k_{\mu(t)} \over
\partial x_i \partial x_j} \right|_{x=x^{\alpha}}.
\end{eqnarray}
The exponent coefficients $q_{ij}^{\alpha}$ remain time dependent
even when the distribution tends to a $\delta$-function. It means
(in this case) that peaks became infinitely narrow. Nevertheless, it
is possible to change variables and represent the weak$^*$ tendency
to stationary discrete distribution as usual tendency to a fixed
points, see (\ref{drifteqlog}) below.

These relations (\ref{selrela}) do not form a closed system of
equations, because the right-hand parts are not functions of
$x_i^{\alpha}$ and $q_{ij}^{\alpha}$. For sufficiently narrow peaks
there should be separation of the relaxation times between the
dynamics {\it on } the support and the  dynamics {\it of } the
support: the relaxation of peak amplitudes (it can be approximated
by the relaxation of the distribution with the finite support,
$\{x^{\alpha}\}$) should be significantly faster than the motion of
the locations of the peaks, the dynamics of $\{x^{\alpha}\}$. Let us
write the first term of the corresponding asymptotics \cite{G1}.

For the finite support $\{x^{\alpha}\}$ the distribution is
$\mu=\sum_{\alpha} N_{\alpha} \delta(x-x^{\alpha})$. Dynamics of
the finite number of variables, $ N_{\alpha}$ obeys the system of
ordinary differential equations
\begin{equation}\label{findrift}
{\D N_{\alpha} \over \D t}=k_{\alpha}({\NN})N_{\alpha},
\end{equation}
where $\NN$ is vector with components $N_{\alpha}$,
$k_{\alpha}({\NN})$ is the value of the reproduction coefficient
$k_{\mu}$ at the point $x^{\alpha}$:
$$k_{\alpha}({\NN})=k_{\mu}(x^{\alpha}) \; \mbox{for} \;
\mu=\sum_{\alpha} N_{\alpha} \delta(x-x^{\alpha}).$$

Let the dynamics of the system (\ref{findrift}) for a given set of
initial conditions be simple: the motion $\NN(t)$ goes to the
stable fixed point $\NN = \NN^*(\{x^{\alpha} \})$. Then we can
take in the right hand side of (\ref{selrela})
\begin{equation}\label{rightpartdri}
\mu(t)=\mu^*(\{x^{\alpha}(t) \}) =\sum_{\alpha}
N^*_{\alpha}\delta(x-x^{\alpha}(t)).
\end{equation}
Because of the time separation we can assume that (i) relaxation
of the amplitudes of peaks is completed and (ii) peaks are
sufficiently narrow, hence, the difference between true
$k_{\mu(t)}$ and the reproduction coefficient for the measure
(\ref{rightpartdri}) with the finite support $\{x^{\alpha}\}$ is
negligible. Let us use the notation $k^*(\{x^{\alpha} \})(x)$ for
this reproduction coefficient. The relations (\ref{selrela})
transform into the ordinary differential equations
\begin{eqnarray}\label{drifteq}
\sum_j q_{ij}^{\alpha} {\D x_j^{\alpha} \over \D t}& =& \left.
{\partial k^*(\{x^{\beta} \})(x) \over
\partial x_i} \right|_{x=x^{\alpha}} ;   \nonumber \\ {\D q_{ij}^{\alpha} \over \D t}& = &
\left.-{\partial^2 k^*(\{x^{\beta} \})(x) \over
\partial x_i \partial x_j} \right|_{x=x^{\alpha}}.
\end{eqnarray}
For many purposes it may be useful to  switch to the logarithmic
time $\tau= \ln t$ and to new variables $$b_{ij}^{\alpha}= {1
\over t} q_{ij}^{\alpha}= \left.-{\partial^2 \langle k(\mu)
\rangle _t \over
\partial x_i \partial x_j}\right|_{x=x^{\alpha}}.$$ For large $t$ we obtain from (\ref{drifteq})
\begin{eqnarray}\label{drifteqlog}
\sum_j b_{ij}^{\alpha} {\D x_j^{\alpha} \over \D \tau}& =& \left.
{\partial k^*(\{x^{\beta} \})(x) \over
\partial x_i} \right|_{x=x^{\alpha}} ;   \nonumber \\ {\D b_{ij}^{\alpha} \over \D \tau}& = &
\left.-{\partial^2 k^*(\{x^{\alpha} \})(x) \over
\partial x_i \partial x_j} \right|_{x=x^{\beta}}- b_{ij}^{\alpha}.
\end{eqnarray}
The way of constructing the drift equations
(\ref{drifteq},\ref{drifteqlog}) for a specific system
(\ref{sel1}) is as follows:
\begin{enumerate}
\item{For finite sets $\{x^{\alpha}\}$ one studies systems
(\ref{findrift}) and finds the equilibrium solutions
$\NN^*(\{x^{\alpha}\})$}; \item{For given measures
$\mu^*(\{x^{\alpha}(t) \})$ (\ref{rightpartdri}) one calculates
the reproduction coefficients $k_{\mu}(x)=k^*(\{x^{\alpha} \})(x)$
and first derivatives of these functions in $x$ at points
$x^{\alpha}$. That is all, the drift equations
(\ref{drifteq},\ref{drifteqlog}) are set up.}
\end{enumerate}

The drift equations (\ref{drifteq},\ref{drifteqlog}) describe the
dynamics of the peaks positions $x^{\alpha} $ and of the
coefficients $q_{ij}^{\alpha}$. For given $x^{\alpha},\:
q_{ij}^{\alpha}$ and $ N^*_{\alpha}$ the distribution density
$\mu$ can be approximated as the sum of narrow Gaussian peaks:
\begin{equation}\label{selGauss}
\mu = \sum_{\alpha} N^*_{\alpha} {\sqrt{\det Q^{\alpha} \over
(2\pi)^{n}}} \exp\left(-{1 \over 2} \sum_{ij}
q_{ij}^{\alpha}(x_i-x^{\alpha}_i)(x_j-x^{\alpha}_j)\right),
\end{equation}
where $Q^{\alpha}$ is the inverse covariance matrix
$(q_{ij}^{\alpha})$.

If the limit dynamics of the system (\ref{findrift})  for finite
supports at $t \rightarrow \infty$ can be described by a more
complicated attractor, then instead of reproduction coefficient
$k^*(\{x^{\alpha} \})(x)=k_{\mu^*}$ for the stationary measures
$\mu^*$ (\ref{rightpartdri}) one can use the average reproduction
coefficient with respect to the corresponding {\it
Sinai--Ruelle--Bowen measure} \cite{Katok,Katok2}. If finite systems
(\ref{findrift}) have several attractors for given $\{x^{\alpha}
\}$, then the dependence $k^*(\{x^{\alpha} \})$ is multi-valued, and
there may be bifurcations and hysteresis with the function
$k^*(\{x^{\alpha} \})$ transition   from one sheet to another. There
are many interesting effects concerning peaks' birth,
desintegration, divergence, and death, and the drift equations
(\ref{drifteq},\ref{drifteqlog}) describe the motion in a
non-critical domain, between these critical effects.

Inheritance (conservation of support) is never absolutely exact.
Small variations, mutations, immigration in biological systems are
very important. Excitation of new degrees of freedom, modes
diffusion, noise are present in physical systems. How does small
perturbation in the inheritance affect the effects of selection?
The answer is usually as follows: there is such a value of
perturbation of the right-hand side of  (\ref{sel1}), at which
they would change nearly nothing, just the limit $\delta$-shaped
peaks transform into sufficiently narrow peaks, and zero limit of
the velocity of their drift at $t \rightarrow \infty$ substitutes
by a small finite one.

\subsection{Drift in presence of mutations and scaling invariance}

The simplest model for ``inheritance + small variability" is given
by a perturbation of (\ref{sel1}) with diffusion term
\begin{equation}\label{selDiff}
{\partial \mu(x,t) \over \partial t}= k_{\mu(x,t)} \times \mu(x,t)
+ \varepsilon \sum_{ij} d_{ij}(x) {\partial ^2 \mu(x,t) \over
\partial x_i \partial x_j}.
\end{equation}
where $\varepsilon>0$ and the matrix of diffusion coefficients
$d_{ij}$ is symmetric and positively definite.

There are almost always no qualitative changes in the asymptotic
behavior, if $\varepsilon$ is sufficiently small. With this the
asymptotics is again described by the drift equations
(\ref{drifteq},\ref{drifteqlog}), modified by taking into account
the diffusion as follows:
\begin{eqnarray}\label{drifteqDif}
\sum_j q_{ij}^{\alpha} {\D x_j^{\alpha} \over \D t}& =& \left.
{\partial k^*(\{x^{\beta} \})(x) \over
\partial x_i} \right|_{x=x^{\alpha}} ;   \nonumber \\ {\D q_{ij}^{\alpha} \over \D t}& = &
\left.-{\partial^2 k^*(\{x^{\beta} \})(x) \over
\partial x_i \partial x_j} \right|_{x=x^{\alpha}} -
2\varepsilon\sum_{kl}q_{ik}^{\alpha}d_{kl}(x^{\alpha})q_{lj}^{\alpha}.
\end{eqnarray}
Now, as distinct from (\ref{drifteq}), the eigenvalues of the
matrices $Q^{\alpha}= (q_{ij}^{\alpha})$ cannot grow infinitely.
This is prevented by the quadratic terms in the right-hand side of
the second equation (\ref{drifteqDif}).

Dynamics of (\ref{drifteqDif}) does not depend  on the value
$\varepsilon>0$ qualitatively, because of the obvious scaling
property.  If $\varepsilon$ is multiplied by a positive number
$\nu$, then, upon rescalling $t' = \nu^{-1/2} t$ and
${q_{ij}^{\alpha}}' =\nu^{-1/2} q_{ij}^{\alpha}$, we have the same
system again. Multiplying $\varepsilon>0$ by $\nu>0$ changes only
peak's velocity values by a factor $\nu^{1/2}$, and their width by
a factor $\nu^{1/4}$. The paths of peaks' motion do not change at
this for the drift approximation (\ref{drifteqDif}) (but the
applicability of this approximation may, of course, change).

\subsection{Three main types of stability}\label{stab}

Stable steady-state solutions of equations of the form
(\ref{sel1}) may be only the sums of $\delta$-functions -- this
was already mentioned. There is a set of specific conditions of
stability, determined by the form of equations.

Consider a stationary distribution for (\ref{sel1}) with a finite
support
$$\mu^*(x)=\sum_{\alpha}N^*_{\alpha} \delta(x-x^{* \alpha}).$$
Steady state of $\mu^*$ means, that
\begin{equation}\label{selstat}
k_{\mu^*}(x^{* \alpha})= 0 \; \mbox{for all} \; \alpha.
\end{equation}

The {\it internal stability} means, that this distribution is stable
with respect to perturbations not increasing the support of $\mu^*$.
That is, the vector $N^*_{\alpha}$ is the stable fixed point for the
dynamical system (\ref{findrift}). Here, as usual, it is possible to
distinguish between the Lyapunov stability, the asymptotic stability
and the first approximation stability (negativeness of real parts
for the eigenvalues of the matrix $\partial \dot{N}^*_{\alpha} /
\partial N^*_{\alpha}$ at the stationary points).

The {\it external stability} ({\it uninvadability}) means
stability to an expansion of the support, i.e. to adding to
$\mu^*$ of a small distribution whose support contains points not
belonging to ${\rm supp} \mu^*$. It makes sense to speak about the
external stability only if there is internal stability. In this
case it is sufficient to restrict ourselves with
$\delta$-functional perturbations. The external stability has a
very transparent physical and biological sense. It is stability
with respect to {\it introduction} into the systems of a new
inherited unit (gene, variety, specie...) in a small amount.

The {\it necessary condition for the external stability} is: the
points $\{ x^{* \alpha}\}$ are points of the global maximum of the
reproduction coefficient $k_{\mu^*}(x)$. It can be formulated as
the optimality principle
\begin{eqnarray} \label{optsel2}
k_{\mu^*}(x)\leq 0 \; \mbox{for all $x$}; \; k_{\mu^*}(x^{*
\alpha})=0.
\end{eqnarray}
The {\it sufficient condition  for the external stability} is: the
points $\{ x^{* \alpha}\}$ and only these points are points of the
global maximum of the reproduction coefficient $k_{\mu^*}(x^{*
\alpha})$. At the same time it is the condition of the external
stability in the first approximation and the optimality principle
\begin{eqnarray} \label{optsel3}
k_{\mu^*}(x) < 0 \; \mbox{for $x \notin \{x^{* \alpha}\}$}; \;
k_{\mu^*}(x^{* \alpha})=0.
\end{eqnarray}
The only  difference from (\ref{optsel2}) is the change of the
inequality sign from $k_{\mu^*}(x)\leq 0 $ to $k_{\mu^*}(x)<0$ for
$x \notin \{x^{* \alpha}\}$. The necessary condition
(\ref{optsel2}) means, that the small $\delta$-functional addition
will not grow in the first approximation. According to the
sufficient condition (\ref{optsel3}) such a small addition will
exponentially decrease.

If $X$ is a finite set, then the combination of the external and the
internal stability is equivalent to the standard stability for a
system of ordinary differential equations.

For the continuous $X$ there is one more kind of stability important
from the applications viewpoint. Substitute $\delta$-shaped peaks at
the points $\{x^{* \alpha}\}$ by narrow Gaussians and shift slightly
the positions of their maxima away from the points $x^{* \alpha}$.
How will the distribution from such initial conditions evolve? If it
tends to $\mu$ without getting too distant from this steady state
distribution, then we can say that the third type of stability --
{\it stable realizability} -- takes place. It is worth mentioning
that the perturbation of this type is only weakly$^*$ small, in
contrast to perturbations considered in the theory of internal and
external stability. Those perturbations are small by their norms.
Let us remind that the norm of the measure $\mu$ is $\|\mu\|=
\sup_{|f| \leq 1} [\mu,f]$. If one shifts the $\delta$-measure of
unite mass by any nonzero distance $\varepsilon$, then the norm of
the perturbation is 2. Nevertheless, this perturbation weakly$^*$
tends to 0 with $\varepsilon \rightarrow 0$.

In order to formalize the condition of stable realizability it is
convenient to use the drift equations in the form
(\ref{drifteqlog}). Let the distribution $\mu^*$ be internally and
externally stable in the first approximations. Let the points
$x^{*\alpha}$ of global maxima of $k_{\mu^*}(x)$ be non-degenerate
in the second approximation. This means that the matrices
\begin{eqnarray}\label{matrseldri}
b_{ij}^{* \alpha}= - \left({\partial ^2 k_{\mu^*}(x) \over
\partial x_i \partial x_j}\right)_{x=x^{*\alpha}}
\end{eqnarray}
are strictly positively definite for all $\alpha$.

Under these conditions of stability and non-degeneracy the
coefficients of (\ref{drifteqlog}) can be easily calculated using
Taylor series expansion in  powers of $(x^{\alpha}-x^{*\alpha})$.
The stable realizability of $\mu^*$ in the first approximation
means that the fixed point of the drift equations
(\ref{drifteqlog}) with the coordinates
\begin{eqnarray}\label{pointsta}
x^{\alpha}=x^{*\alpha}, \; b_{ij}^{\alpha}=b_{ij}^{* \alpha}
\end{eqnarray}
is stable in the first approximation. It is the usual stability
for the system (\ref{drifteqlog}) of ordinary differential
equations, and these conditions with the notion of the stable
realizability  became clear from the logarithmic time drift
equations (\ref{drifteqlog}) directly.

To explain the sense of the stable realizability we used in the
book \cite{GorKhleb} the idea of the ``Gardens of Eden" from
J.H.Conway ``Game of Life" \cite{GardGard}. That are Game of Life
patterns which have no father patterns and therefore can occur
only at generation 0, from the very beginning. It is not known if
a pattern which has a father pattern, but no grandfather pattern
exists. It is the same situation, as for internal and external
stable (uninvadable) state which is not stable realizable: it
cannot be destroyed by mutants invasion and by the small variation
of conditions, but, at the same time, it is not attractive for
drift, and, hence, can not be realized in this asymptotic motion.
It can be only created.

The idea of drift and the corresponding stability notions become
necessary in any approach to evolutionary dynamics on continuous
paces. In recent paper \cite{EshSan}, the asymptotic stability
under the replicator dynamics over a continuum of pure strategies
was studied. It was shown in \cite{EshSan} that strong
uninvadability of a pure strategy $x^*$ \cite{Bom90} is
insufficient for its stability with respect to the drift: It does
not imply convergence to $x^*$ when starting from a distribution
of small deviations from $x^*$, regardless of how small these
deviations are. The standard idea of asymptotic stability is:
``after small deviation the system returns to the initial regime,
and do not deviate to much on the way of returning".  The crucial
question for the measure dynamics is: in which topology the
deviation is small? The small shift of the narrow peak of
distribution in the continuous space of strategies can be
considered as a small deviation in the weak$^*$ topology, but it
is definitely large deviation in the strong topology, for example,
if the shift is not small in comparison with the peak with. In the
papers \cite{Oechssler01,Oechssler02} the idea of drift equations
appeared again for the gaussian peaks in the dynamics of
continuous symmetric evolutionary games. The authors
\cite{Oechssler01,Oechssler02} introduced the idea of
``evolutionary robustness" (realizability) and claimed  the
necessity of the additional notion of stability very
energetically: ``Furthermore, we provide new conditions for the
stability of rest points and show that even strict equilibria may
be unstable".

\section{Example: Cell division
self-synchronization}\label{cellSinch}

The results described above admit for a whole family of
generalizations. In particular,  it seems to be important to
extend the theorems of selection to the case of vector
distributions, when $k_{\mu}(x)$ is a linear operator at each
$\mu, x$. In this case, in the optimality principles for
steady-states distributions  the maximal eigenvalues of these
operators $k_{\mu}(x)$ appear instead of the values of the
reproduction coefficients. For general $\omega$-limit sets special
multiplicative operator averages are in use \cite{G1}. It is
possible also to make generalizations for some classes of
non-autonomous equations with explicit dependencies of
$k_{\mu}(x)$ on $t$ \cite{G1}.

Availability of such a network of generalizations allows to
construct the reasoning as follows: {\it what} is inherited (i.e.
for what the law of conservation of support holds) is the subject
of selection (i.e. with respect to these variables at $t
\rightarrow \infty$ the distribution becomes discrete and the
limit support can be described by the optimality principles).

This section gives a somewhat unconventional example of
inheritance and selection, when the reproduction coefficients are
subject additional conditions of symmetry.

Consider a culture of microorganisms in a certain medium (for
example, pathogenous microbes in the organism of a host). Assume,
for simplicity, the following: {\it let the time period spent by
these microorganisms for the whole life cycle be identical}.

At the end of the life cycle the microorganism disappears and new
several microorganisms appear in the initial phase. Let $T$ be the
time of the life cycle. Each microorganism holds the value of the
inherited variable, it is ``the moment of its appearance (${\rm
mod} T$)". Indeed, if the given microorganism emerges at time
$\tau$ ($0< \tau \leq T$), then its first descendants appear at
time $T + \tau$, the next generation -- at the moment $2T + \tau$,
then $3T + \tau$ and so on.

It is natural to assume that the phase $\tau$ (${\rm mod} T$) is the
inherited variable with some accuracy. This implies selection of
phases and, therefore, survival of their discrete number $\tau_1,
\ldots \tau_m$, only. But results of the preceding sections cannot
be applied directly to this problem. The reason is the additional
symmetry of the system with respect to the phase shift. But the
typicalness of selection and the instability of the uniform
distribution over the phases $\tau$ (${\rm mod} T$) can be shown for
this case, too. Let us demonstrate it with the simplest model.

Let the difference between the microorganisms at each time moment
be related to the difference in the development phases only. Let
us also assume that the state of the medium can be considered as a
function of the distribution $\mu(\tau)$ of microorganisms over
the phases $\tau \in ]0,T]$ (the quasi-steady state approximation
for the medium). Consider the system at discrete times $nT$ and
assume the coefficient connecting $\mu$ at moments $nT$ and $nT+T$
to be the exponent of the linear integral operator value:
\begin{equation}\label{sel20}
\mu_{n+1}(\tau)=\mu_{n}(\tau) \exp \left[k_0 - \int_0^T k_1(\tau -
\tau') \mu_n(\tau')\, \D \tau'\right].
\end{equation}
Here, $\mu_n(\tau)$ is the distribution at the moment $nT$,
$k_0={\rm const}$, $k_1(\tau)$ is a periodic function of period
$T$.

This model is constructed in order to study the interaction of two
factors of microbial dynamics: the fixed period of the cell cycle,
and the density--dependent interaction with the medium. The
density--dependent interactions is modeled in the same degree of
generality, as in general Volterra equations, with  one addition:
for systems with discrete time the exponential form of
reproduction coefficient is more natural and useful, it was shown
by Ricker \cite{Ricker}. In this sense, (\ref{sel20}) presents a
hybrid Volterra--Ricker model for microbial population with the
fixed  period of the cell cycle. The deviation from the fixed
period can be formalized as a phase diffusion, as it was presented
for the general systems with inheritance in the previous section,
and here we study the ``pure" consequences of  phase inheritance.

This model significantly differs from the continuous time models,
where the cell splitting is presented as a ``quasi-chemical
process" of fission of a cell with size parameter $2x$ into two
identical daughters with parameter $x$: $[2x] \rightarrow [x]
+[x]$ (any cell with size parameter $x$ can spontaneously split at
any time without any dependence on its history). The probability
of splitting depends here on the cell size only.

For example, a linear model for the growth of such a
size-structured cell population, reproducing by continuous
``Markov" fission of cells  is formulated and identified in
\cite{DiekCell}. With known functions $\alpha(x)$ (death), $g(x)$
(growth) and $b(x)$ (``loss due to splitting") the model takes the
form of a first order partial differential equation,
\begin{equation}\label{DiekC}
\frac{\partial n(t,x)}{\partial t} +
\frac{\partial(gn(t,x))}{\partial x}=-\alpha
n(t,x)-bn(t,x)+4b(2x)n(t,2x),
\end{equation}
for the density $n(x,t)$. The equation is accompanied by proper
initial and boundary conditions. With several restrictions it is
proved that the asymptotic behavior of solutions for $t\rightarrow
\infty$ has the form $n(t,x)= ce^ {\lambda_d
t}(\tilde{n}(x)+o(1))$, where the constant $c$ (single in the
expression) depends on the initial distribution. The way to
establish the sign of $\lambda_d$ and the explicit form of the
function $\tilde{n}(x)$ is indicated. The continuous time Markov
property (independence of history) with continuous kinetic
coefficient $b(x)$ implies smoothing of limit distributions. In
model (\ref{sel20}) the cell remembers the moment of its ``birth"
and splits exactly after living time $T$. This property is the
main difference between (\ref{sel20}) and (\ref{DiekC}). The
second difference is nonlinearity of (\ref{sel20}), we take into
account the cells density--dependent interaction (mediated by the
medium state). The coefficient of this interaction, $k_1(\tau -
\tau')$, depends on the age difference of interacting cells. This
nonlinearity in (\ref{sel20}) includes in implicit form the
structure of population with age--determined size difference, etc.

The uniform steady-state  $\mu^* \equiv n^* = {\rm const}$ for
(\ref{sel20}) is:
\begin{equation}\label{selOdnor}
n^* = {k_0 \over \int_0^T k_1(\theta)\, \D \theta}.
\end{equation}

In order to examine stability of the uniform steady state $\mu^*$
(\ref{selOdnor}), the system (\ref{sel20}) is linearized. For
small deviations $\Delta \mu(\tau)$ in linear approximation
\begin{equation}\label{sel21}
\Delta \mu_{n+1}(\tau) = \Delta \mu_{n}(\tau) - n^* \int_0^T
k_1(\tau - \tau')\Delta \mu_n(\tau')\, \D \tau'.
\end{equation}

Expand $k_1(\theta)$ into the Fourier series:
\begin{equation}\label{selFour}
k_1(\theta)=b_0 + \sum_{n=1}^\infty \left(a_n \sin \left(2 \pi n
{\theta \over T}\right) + b_n \cos \left(2 \pi n {\theta \over
T}\right) \right).
\end{equation}

Denote by $A$ operator of the right-hand side of (\ref{sel21}). In
the basis of functions $$e_{{\rm s}\, n }=\sin\left(2 \pi n
{\theta \over T}\right), \; e_{{\rm c}\, n }=\cos\left(2 \pi n
{\theta \over T}\right)$$ on the segment $]0,T]$ the operator $A$
is block-diagonal. The vector $e_0$ is eigenvector,
$Ae_0=\lambda_0 e_0$, $\lambda_0=1-n^* b_0T$. On the
two-dimensional space, generated by vectors $e_{{\rm s}\, n },
\,e_{{\rm c}\, n }$ the operator $A$ is acting as a matrix
\begin{eqnarray}\label{sel22}
A_n = \left( \begin{array}{cc} 1- {Tn^* \over 2}b_n \, & - {T
n^*\over 2}a_n \\ {Tn^* \over 2}a_n \, & \, 1- {Tn^* \over 2}b_n
\end{array} \right).
\end{eqnarray}
The corresponding eigenvalues are
\begin{equation}\label{selEigen}
\lambda_{n \, 1,2}=1-{Tn^* \over 2}(b_n \pm i a_n).
\end{equation}

For the uniform steady state $\mu^*$ (\ref{selOdnor}) to be
unstable it is sufficient that the absolute value of at least one
eigenvalue $\lambda_{n \, 1,2}$ be larger than 1: $|\lambda_{n \,
1,2}|>1$. If there is at least one negative  Fourier
cosine-coefficient $b_n<0$, then ${\rm Re}\lambda_n > 1 $, and
thus $|\lambda_n|>1$.

Note now, that almost all periodic functions (continuous, smooth,
analytical -- this does not matter) have negative Fourier
cosine-coefficient. This can be understood as follows. The
sequence $b_n$ tends to zero at $n \rightarrow \infty$. Therefore,
if all $b_n\geq 0$, then, by changing $b_n$ at sufficiently large
$n$, we can make $b_n$ negative, and the perturbation value can be
chosen less than any previously set positive number. On the other
hand, if some $b_n<0$, then this coefficient cannot be made
non-negative by sufficiently small perturbations. Moreover, the
set of functions that have all Fourier cosine-coefficient
non-negative is completely thin, because for any compact of
functions $K$ (for most of norms in use) the sequence $B_n =
\max_{f \in K}{|b_n (f)|}$ tends to zero, where $b_n (f)$ is the
$n$th Fourier cosine-coefficient of function $f$.

The model (\ref{sel20}) is revealing, because for it we can trace
the dynamics over large times, if we restrict ourselves with a
finite segment of the Fourier series for $k_1(\theta)$. Describe
it for
\begin{equation}\label{selFour00}
k_1(\theta)=b_0 + a \sin \left(2 \pi  {\theta \over T}\right) + b
\cos \left(2 \pi  {\theta \over T}\right).
\end{equation}
Assume further that  $b<0$ (then the homogeneous distribution
$\mu^*\equiv {k_0 \over b_0 T}$ is unstable) and $b_0 >\sqrt{
a^2+b^2}$ (then the $ \int \mu(\tau) \, \D \tau$ cannot grow
unbounded in time). Introduce notations
\begin{eqnarray}
&&M_0(\mu)= \int_0^T \mu(\tau) \, \D \tau, \: M_{\rm c}(\mu)=
\int_0^T \cos \left(2 \pi  {\tau \over T}\right) \mu(\tau) \, \D
\tau, \nonumber \\&& M_{\rm s}(\mu)= \int_0^T \sin \left(2 \pi
{\tau \over T}\right) \mu(\tau) \, \D \tau, \: \langle \mu
\rangle_n = {1 \over n}\sum_{m=0}^{n-1} \mu_m,
\end{eqnarray}
where $\mu_m$ is the distribution $\mu$ at the discrete time $m$.

In these notations,
\begin{eqnarray}
\mu_{n+1}(\tau)&=& \mu_{n}(\tau) \exp\left[k_0-b_0
M_0(\mu_n)-(aM_{\rm c}(\mu_n)+bM_{\rm s}(\mu_n))\sin \left(2 \pi
{\tau \over T}\right) \right. \nonumber \\ &&\left. +(aM_{\rm
s}(\mu_n)-bM_{\rm c}(\mu_n))\cos \left(2 \pi {\tau \over
T}\right)\right].
\end{eqnarray}
Represent the distribution $\mu_{n}(\tau)$ through the initial
distribution $\mu_{0}(\tau)$ and the functionals $M_0, M_{\rm c},
M_{\rm s}$ values for the average distribution
$\langle\mu\rangle_n)$:
\begin{eqnarray}\label{sel25}
\mu_{n}(\tau)&=& \mu_{0}(\tau)\nonumber \\ &&\times \exp\left\{ n
\left[k_0-b_0 M_0(\langle\mu\rangle_n)-(aM_{\rm
c}(\langle\mu\rangle_n)+bM_{\rm s}(\langle\mu\rangle_n))\sin
\left(2 \pi  {\tau \over T}\right) \right. \right. \nonumber
\\ &&\left.\left. +(aM_{\rm s}(\langle\mu\rangle_n)-
bM_{\rm c}(\langle\mu\rangle_n))\cos \left(2 \pi {\tau \over
T}\right)\right]\right\}.
\end{eqnarray}

The exponent  in (\ref{sel25}) is either  independent of $\tau$,
or there is a function with the single maximum on $]0,T]$. The
coordinate $\tau_n^\#$ of this maximum is easily calculated
\begin{eqnarray}\label{sel26}
\tau_n^\# = - {T \over 2 \pi}\arctan{aM_{\rm
c}(\langle\mu\rangle_n)+bM_{\rm s}(\langle\mu\rangle_n) \over
aM_{\rm s}(\langle\mu\rangle_n)- bM_{\rm c}(\langle\mu\rangle_n)}
\end{eqnarray}

Let the non-uniform smooth initial distribution $\mu_{0}$ has the
whole segment $]0,T]$ as its support. At the time progress the
distributions $\mu_n(\tau)$ takes the shape of ever narrowing peak.
With high accuracy at large $a$ we can approximate $\mu_n(\tau)$ by
the Gaussian distribution (approximation accuracy is understood in
the weak$^*$ sense, as closeness of mean values):
\begin{eqnarray}\label{selSinGa}
&&\mu_n(\tau) \approx M_0 \sqrt{{q_n \over \pi}} \exp [- q_n
(\tau-\tau_n^\# )^2], \, M_0={k_0 \over k_1(0)}= {k_0 \over b_0+b},
\\ &&q_n^2=n^2\left({2 \pi \over T}\right)^4 \nonumber \\  && \; \; \; \times  \left[(aM_{\rm
c}(\langle\mu\rangle_n)+bM_{\rm s}(\langle\mu\rangle_n))^2 +
(aM_{\rm s}(\langle\mu\rangle_n)- bM_{\rm
c}(\langle\mu\rangle_n))^2\right].\nonumber
\end{eqnarray}
Expression (\ref{selSinGa}) involves the average measure
$\langle\mu\rangle_n$ which is difficult to compute. However, we
can operate without direct computation of $\langle\mu\rangle_n$.
At $q_n\gg {1 \over T^2 }$ we can compute $q_{n+1}$ and
$\tau_{n+1}^\#$:
\begin{eqnarray}\label{sel28}
&&\mu_{n+1} \approx M_0 \sqrt{{q_n+ \Delta q \over \pi}} \exp
\left[-(q_n+ \Delta q)((\tau-\tau_n^\# - \Delta \tau^\#)^2
\right], \nonumber \\ &&\Delta q \approx - {1\over 2}bM_0 \left({2
\pi \over T}^2\right), \: \Delta \tau^\# \approx {1 \over q}M_0 {2
\pi \over T}.
\end{eqnarray}

The accuracy of these expression grows with time $n$. The value
$q_n$ grows at large $n$ almost linearly, and $\tau_n^\#$,
respectively, as the sum of the harmonic series $({\rm mod}T)$,
i.e. as $\ln n$ $({\rm mod}T)$. The drift effect takes place:
location of the peak $\tau_n^\#$, passes at $n \rightarrow \infty$
the distance diverging as $\ln n$.

Of interest is the case, when $b > 0$ but $$|\lambda
_1|^2=\left(1+n^*b{T \over 2}\right)^2+\left(n^*a{T \over
2}\right)^2>1.$$ With this, homogeneous distribution $\mu^*\equiv
n^*$ is not stable but $\mu$ does not tend to $\delta$-functions.
There are smooth stable ``self-synchronization waves" of the form
$$\mu_n=\gamma \exp\left[q \cos\left((\tau-n\Delta \tau^\#){2\pi
\over T}\right) \right].$$ At small $b>0$ ($b\ll |a|$, $bM_0\ll
a^2$)  we can find explicit form of approximated expressions for
$q$ and $\Delta \tau^\#$:
\begin{equation}
q\approx {a^2 M_0 \over 2b}, \; \Delta \tau^ \# \approx {b T \over
\pi a}.
\end{equation}

At $b>0, b\rightarrow 0$, smooth self-synchronization waves become
ever narrowing peaks, and their steady velocity approaches zero. If
$b=0, \, |\lambda_1|^2>1$, then the effect of selection takes place
again, and for almost all initial conditions $\mu_0$ with the
support being the whole segment $]0,T]$ the distribution $\mu_n$
takes at large $n$ the form of a slowly drifting almost Gaussian
peak. It becomes narrower with the time, and the motion slows down.
Instead of the linear growth of $q_n$ which takes place at $b<0$
(\ref{sel28}), for $b=0$,  $q_{n+1}-q_n\approx {\rm const }q_n^{-1}$
and $q_n$ grows as ${\rm const} \sqrt{n}$.

\begin{figure}[t]
\centering {
\includegraphics[width=90mm, height=80mm]{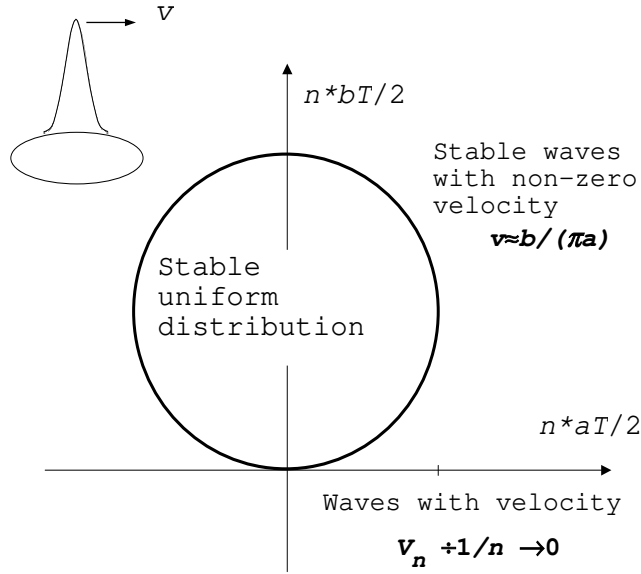}
\caption {\label{AutoSinch}The simplest model of cell division
self-synchronization:  The parametric portrait.}}
\end{figure}

The parametric portrait of the system for the simple reproduction
coefficient (\ref{selFour00}) is presented in Fig.
\ref{AutoSinch}.

As it is usual, a small desynchronization transforms
$\delta$-functional limit peaks to narrow Gaussian peaks, and the
velocity of peaks tends to small but nonzero velocity instead of
zero. The systems with small desynchronization  can be described
by equations of the form (\ref{drifteqDif}). The large
desynchronization can completely destroy the effects of phase
selection and, for example, it might lead to the globally stable
uniform phase distribution.

There are many specific mechanisms of synchronization and
desynchronysation in physics and biology (see, for example,
\cite{Syn0,Syn1,Syn2,Syn3,Syn4}). We described here very simple, but
universal mechanism: it requires only that the time of the life
cycle is fixed, in this case in a generic situation we should
observe the self-synchronization. Of course, the real-world
situation can be much more complicated, with a plenty of additional
factors, but the basic mechanism of the ``phase selection" works
always, if the life cycle has more or less fixed duration.

\section{Conclusion: Main results about systems with inheritance}\label{main}

\begin{enumerate}
\item{If a kinetic equation has the {\it quasi-biological form}
(\ref{sel1}), then it has a rich system of invariant manifolds:
for any closed subset $A \subset X$ the set of distributions
$\MM_A= \{\mu \, \mid \, {\rm supp} \mu \subseteq A \}$ is
invariant with respect to the system (\ref{sel1}). These invariant
manifolds form important algebraic structure, the summation of
manifolds is possible: $$\MM_A\oplus \MM_B = \MM_{A \cup B}.$$ (Of
course, $\MM_{A \cap B}= \MM_A \cap \MM_B$).}
\item{Typically, all the $\omega$-limit points belong to invariant
manifolds $\MM_A$ with finite $A$ (from the application point of
view there is no difference between finite and almost finite
sets). The finite-dimensional approximations of the reproduction
coefficient (\ref{sel7}) provides the minimax estimation of the
number of points in $A$.}
\item{Typically, systems with inheritance have a rich family of
Lyapunov functionals of form (\ref{entropiika}) similar to the
Burg entropy, these functionals can be interpreted as the measures
of decreasing diversity.}
\item{For systems with inheritance (\ref{sel1}) a solution
typically  tends to be a finite set of narrow peaks getting in time
more and more narrow, moving slower and slower. It is possible that
these peaks do not tend to fixed positions, rather they continue
moving, and the path covered tends to infinity at $t \rightarrow
\infty$. This is the {\it drift effect.}} \item{The equations for
peak dynamics, the drift equations,
(\ref{drifteq},\ref{drifteqlog},\ref{drifteqDif}) describe dynamics
of the shapes of the peaks and their positions. For systems with
small variability (``mutations") the drift equations
(\ref{drifteqDif}) has the scaling property: the change of the
intensity of mutations is equivalent to the change of the time
scale.} \item{Three specific types of stability are important for
the systems with inheritance: internal stability (stability with
respect to perturbations without extension of distribution support),
external stability (stability with respect to small one-point
extension of distribution support), and stable realizability
(stability with respect to weakly$^*$ small perturbations: small
extensions and small shifts of the peaks; these perturbations are
small in the weak$^*$ topology.).}
\end{enumerate}

The cell division self-synchronization demonstrates effects of
unusual inherited unit, it is an example of a ``phase selection".
One specific property of this selection is additional symmetry
with respect to phase shift. In this case, the general results
about selection cannot be used directly. Nevertheless, the
``equivariant" selection theory successfully works too.

Some exact results of the mathematical selection theory can be
found in \cite{Kuz1,Kuz2}. There exist many physical examples of
systems with inheritance
\cite{Zakharov1,Zakharov2,Lvov,MOdSel1,MOdSel2,MOdSel3,MOdSel4}. A
wide field of ecological applications was described in the book
\cite{SemSem}. An introduction into adaptive dynamics was given in
notes \cite{Odo} that illustrate largely by way of examples, how
standard ecological models can be put into an evolutionary
perspective in order to gain insight in the role of natural
selection in shaping life history characteristics.

{\bf Acknowledgement.} Author is grateful to  V. Okhonin, who
involved him many years ago in the analysis of mathematical models
of natural selection. M. Gromov kindly allowed to quote his talk
\cite{GroGo}.

\end{document}